\documentclass[12pt,journal,draftcls,a4paper,onecolumn]{IEEEtran} 
\usepackage{dsfont}

\usepackage{subeqnarray}
\usepackage{cleveref}
\usepackage{amssymb}
\usepackage{amsfonts}
\usepackage{amsmath}
\usepackage[dvipdfm]{graphicx}
\usepackage{bmpsize}
\usepackage{amsmath}
\usepackage[all,poly]{xy}
\usepackage{multirow}
\usepackage{algorithm}
\usepackage{algorithmic}
\usepackage{color}
\usepackage[noadjust]{cite}

\title{Robust Bayesian target detection algorithm for depth imaging from sparse single-photon
data}
\author{Yoann Altmann, Ximing Ren, Aongus McCarthy, Gerald S. Buller and Steve
McLaughlin 
\thanks{Yoann Altmann, Ximing Ren, Aongus McCarthy, Gerald Buller and Steve
McLaughlin are with School of Engineering and Physical Sciences, Heriot-Watt University,
U.K. (email: \{Y.Altmann;X.Ren;A.McCarthy; G.S.Buller;S.McLaughlin\}@hw.ac.uk).}
\thanks{This study was supported by EPSRC via grants EP/J015180/1, EP/K015338/1 and EP/M01326X/1.
}}
\newcommand{\bGam}{\boldsymbol{\Gamma}}

\newcommand{\bpsi}{{\boldsymbol \psi}}
\newcommand{\bPhi}{{\boldsymbol \Phi}}
\newcommand{\bTheta}{{\boldsymbol \Theta}}
\newcommand{\btheta}{{\boldsymbol \theta}}
\newcommand{\argmax}{\operatornamewithlimits{argmax}}
\newcommand{\blue}{\textcolor{blue} }
\graphicspath{{figures/}}

\def\bfB{{\mathbf{B}}}

\def\bfT{{\mathbf{T}}}

\def\bfZ{{\mathbf{Z}}}

\def\bbR{{\mathbb{R}}}

\def\bbT{{\mathbb{T}}}

\def\bbZ{{\mathbb{Z}}}

\newcommand{\Vpix}[1]{\mathbf{y}_{#1}}

\newcommand{\MATpix}{\mathbf{Y}}
\newcommand{\pix}[2]{y_{#1,#2}}

\newcommand{\nbbin}{T}
\newcommand{\nobin}{t}

\newcommand{\transp}{^T}

\newcommand{\norm}[1]{\left\|#1\right\|}

\newcommand{\Indicfun}[2]{\textbf{1}_{#1}\left(#2\right)}

\newenvironment{algogo}[1]{
\smallskip
\noindent \hrule\vspace{0.2\baselineskip} \hrule
\begin{small}
\refstepcounter{algo} \center{\bf \textsc{Algorithm \thealgo}}
\\{\center{\bf #1}}
\smallskip
\flushleft
 } {
\end{small}
\smallskip
\hrule\vspace{0.2\baselineskip} \hrule
\smallskip
}

\newcounter{algo}
\renewcommand{\thealgo}{\arabic{algo}}

\begin{document}
\maketitle

\begin{abstract}
This paper presents a new Bayesian model and associated algorithm for depth 
and intensity profiling using full waveforms from time-correlated 
single-photon counting (TCSPC) measurements in the limit of very low photon counts (i.e., typically less than $20$ photons per pixel). The
model represents each Lidar waveform as an unknown constant background level, which is 
combined in the presence of a target, to a known impulse response weighted by the target 
intensity and finally corrupted by Poisson noise. The joint target detection and depth imaging 
problem is expressed as a pixel-wise model selection and estimation problem which is solved 
using Bayesian inference. Prior
knowledge about the problem is embedded in a hierarchical
model that describes the dependence structure between the model
parameters while accounting for their constraints. In particular, Markov
random fields (MRFs) are used to model the joint distribution of the
background levels and of the target presence labels, which are both expected to exhibit significant spatial
correlations. An adaptive Markov chain Monte Carlo algorithm including reversible-jump updates is
then proposed to compute the Bayesian estimates of interest. This algorithm is equipped with a
stochastic optimization adaptation mechanism that automatically
adjusts the parameters of the MRFs by
maximum marginal likelihood estimation. Finally, the benefits of the proposed
methodology are demonstrated through a series of experiments
using real data.
\end{abstract}


\section{Introduction}
Time-of-flight laser detection and ranging (Lidar) based imaging systems are used to reconstruct 3-dimensional scenes in many applications, including automotive \cite{Ogawa2006,Lindner2009,Matzka2012,Yeonsik2012}, environmental sciences \cite{Ramirez2012,Hakala2012}, architectural engineering and defence \cite{Cadalli2002,Gao2011} applications. This challenging problem consists of illuminating the scene with a train of laser pulses and analysing the distribution of the photons received by the detector to infer the presence of objects as well as their range, and radiative properties (e.g., reflectivity, observation conditions,\ldots). Using scanning systems, a histogram of time delays between the emitted pulses and the detected photon arrivals is usually recorded for each pixel, associated with a different region of the scene. Conventionally, in the presence of objects, the recorded photon histograms are decomposed into a series of peaks whose positions can be used to infer the distance of the objects present in each region of the scene and whose amplitudes provide information about the intensity of the objects. 

In this paper, we investigate the target detection problem which consists of inferring the regions or pixels of the scene where objects are present. Moreover, we propose an algorithm for applications where the flux of detected photons is small and for which classical depth imaging methods \cite{McCarthy2013} usually provide unsatisfactory results in terms of range and intensity estimation. This is typically the case for free-space depth profiling on targets at very long distances based on the time-correlated single-photon counting (TCSPC) technique \cite{McCarthy2013}, which negotiates the trade-offs between range/intensity estimation quality, data acquisition time and output laser power. In addition, this might be extended for sparse single-photon depth imaging in turbid media, e.g., underwater depth imaging \cite{Maccarone2015}.
In contrast with the method proposed in \cite{Kirmani2014}, we consider scene observation using a scanning system whose acquisition time per pixel is fixed, thus leading to a deterministic and user-defined overall acquisition duration. As in \cite{Altmann2016a}, the number of detected photons can thus vary across the image pixels and some pixels can be empty (i.e., no detected photons).

In this work, we assume that the targets potentially present in the scene of interest are opaque, i.e., are composed of a single surface per pixel. As in \cite{Kirmani2014,Altmann2016a}, we consider the potential presence of two kinds of detector events: the photons originating from the illumination laser and scattered back from the target (if present); and the background detector events originating from ambient light and the "dark" events resulting from detector noise. The proposed method aims to estimate the respective contributions of the actual target (if any) and the background in the photon timing histograms. 

Following a classical Bayesian approach, as in \cite{Hernandez2007,Wallace2014,Altmann2016a}, we express the target detection and identification problem as a pixel-wise model selection and estimation problem. More precisely, two observation models, conditioned on the presence or absence of a target (modelled by binary labels) are considered for each pixel. We then assign prior distributions to each of the unknown parameters in each model to include available information within the estimation procedure. The probabilities of target presence (or equivalently the binary labels associated with the presence/absence of target) are also assigned prior distributions accounting for spatial organization of the objects in the scene.

Classical Bayesian estimators associated with the joint posterior cannot be
easily computed due to the complexity of the model, in particular because the number of underlying parameters (number of targets) is unknown and potentially large. To tackle this problem, a Reversible-Jump Markov chain Monte Carlo (RJ-MCMC) \cite{Green1995,Andrieu1999} method is
used to generate samples according to this posterior by allowing moves between different parameter spaces. More precisely, we construct an efficient
stochastic gradient MCMC (SGMCMC) algorithm \cite{Pereyra2014ssp} that simultaneously estimates the background levels and the target distances and intensity, along with the MRFs parameters.\\
The main contributions of this work are threefold:
\begin{enumerate}
\item  We develop a new Bayesian algorithm for joint target detection and identification, which takes spatial
correlations affecting the background levels and the target locations into account through Markovian dependencies. To the best of our knowledge, the proposed method is the first joint target detection and identification method designed for depth imaging using single-photon data.
\item  An adaptive Markov chain Monte Carlo algorithm including Reversible-jump updates is proposed to compute the
Bayesian estimates of interest and perform Bayesian inference. This algorithm included RJ-MCMC updates and is equipped
with a stochastic optimization mechanism that adjusts automatically the
parameters of the Markov random fields by maximum marginal likelihood estimation,
thus removing the need to set the regularization parameters, e.g., by cross-validation.
\item We show the benefits of the proposed flexible model for reconstructing a real 3D object in scenarios where the number of detected photons is very low and the background levels are significant. 
\end{enumerate}

The remainder of this paper in organized as follows. Section \ref{sec:model} recalls the statistical models used for depth imaging using time-of-flight scanning sensors, based on TCSPC. Section \ref{sec:Bay_model} presents the new hierarchical Bayesian model which takes into account the inherent spatial correlations between parameters of spatially close pixels. Section \ref{sec:Gibbs} discusses the estimation of the model parameters including the detection labels using adaptive MCMC methods including reversible jump updates. Simulation results conducted using an actual time-of-flight scanning sensor are presented and discussed in Section \ref{sec:simulations}. Finally, conclusions and potential future work are reported in Section \ref{sec:conclusion}.

\section{Problem formulation}
\label{sec:model} 
Assessing the presence of targets from TCSPC measurements in an unsupervised manner is a challenging problem as the detection performance highly depends on the nature of the potential targets (range and reflectivity), as well as the observation conditions (system performance, ambient illumination). In practice, the target range can be restrained to a bounded interval and the estimated reflectivity parameter can be used to assess the target presence (i.e., via thresholding). In a similar manner to \cite{Kirmani2014,Shin2014}, it is possible to enhance the estimation performance of depth and reflectivity parameters by processing simultaneously several pixels/spatial locations, thus improving the subsequent target detection. However, the nature of such sequential processes implies sub-optimal detection performance (which depends on the quality of the previous estimation steps). For this reason, we propose a Bayesian model and method allowing the joint target detection and depth/reflectivity estimation. In contrast to \cite{Kirmani2014,Shin2014}, in this work we assume that the ambient noise level, which can vary among pixels, is unknown and thus needs to be estimated. This is typically the case for long range measurements where the background levels can change due to time-varying illumination conditions. In this work, the background levels are estimated from signal measured during the detection process, and not beforehand, as may be done with a detector array \cite{Gariepy2015}.

We consider a set of $N_{\textrm{row}} \times N_{\textrm{col}}$ observed Lidar waveforms/pixels 
$\Vpix{i,j} = [\pix{i,j}{1},\ldots,\pix{i,j}{\nbbin}]\transp, (i,j) \in \{1,\ldots,N_{\textrm{row}}\} \times \{1,\ldots,N_{\textrm{col}}\}$ where $\nbbin$ is the number of temporal
(corresponding to range) bins.
To be precise, $\pix{i,j}{\nobin}$ is the photon count within the $\nobin$th bin of the pixel or location $(i,j)$. Let $z_{i,j}\in \left\lbrace 0,1 \right\rbrace$ be a binary variable associated with the presence ($z_{i,j}=1$) or absence ($z_{i,j}=0$) of target in the pixel $(i,j)$. In the absence of target, $\pix{i,j}{\nobin}$ is assumed to be drawn from the following Poisson distribution 
\begin{eqnarray}
\label{eq:model0}
\pix{i,j}{\nobin}\left|\left(z_{i,j}=0, \btheta_{i,j}^{0}\right)\sim \mathcal{P}\left(b_{i,j}\right)\right.,
\end{eqnarray}
where $b_{i,j}>0$ stands for the background and dark photon level, which is assumed to be constant in all bins of a given pixel. The model \eqref{eq:model0} for each pixel is denoted $\mathcal{M}_0^{(i,j)}$ ($0$ denotes the absence of a target). Moreover $\btheta_{i,j}^{0}$ denotes the set of likelihood parameters in pixel $(i,j)$ under $\mathcal{M}_0^{(i,j)}$, i.e., $\btheta_{i,j}^{0}=b_{i,j} \in \bbR^+$.

Assume now the presence of a target in the pixel $(i,j)$. Let $\nobin_{i,j}$ be the position of that object surface at a given range from the sensor and $r_{i,j}$ its intensity. The target position is considered as a discrete variable defined on $\bbT=\{t_{min},\ldots,t_{max}\}$, such that $1\leq t_{min}\leq t_{max} \leq T$ (in this paper we set $(t_{min},t_{max})=(1,T)$). In that case, we obtain 
\begin{eqnarray}
\label{eq:model1}
\pix{i,j}{\nobin}\left| \left(z_{i,j}=1, \btheta_{i,j}^{1}\right) \sim \mathcal{P}\left(r_{i,j} g_{0}\left(\nobin-\nobin_{i,j}\right) + b_{i,j}\right)\right.,
\end{eqnarray}
where $g_{0}(\cdot)>0$ is the photon impulse response, which is assumed to be known (this response can be estimated during the imaging system calibration). In a similar fashion to \eqref{eq:model0}, the model \eqref{eq:model1} for each pixel is denoted $\mathcal{M}_1^{(i,j)}$ and in a similar fashion to the background level, the target intensity in each pixel is non-negative, i.e., $r_{i,j} \geq 0$. In \eqref{eq:model1}, $\btheta_{i,j}^{1}$ stands for the set of likelihood parameters in pixel $(i,j)$ under $\mathcal{M}_1^{(i,j)}$, i.e., $\btheta_{i,j}^{1}=[r_{i,j}, t_{i,j}, b_{i,j}] \in \bbR^+ \times \bbT \times \bbR^+$. Note that the background levels in \eqref{eq:model0} and \eqref{eq:model1} have the same physical meaning and are assumed to present the same statistical properties under $\mathcal{M}_0^{(i,j)}$ and $\mathcal{M}_1^{(i,j)}$. Thus, a single background level $b_{i,j}$, independent from the observation model, is used for each pixel. 

The problem addressed in this paper consists of deciding whether a target is present ($\mathcal{M}_0^{(i,j)}$) or not ($\mathcal{M}_1^{(i,j)}$) in each pixel and of estimating the position and intensity of the targets present in the scene, from the observed data gathered in the $N_{\textrm{row}} \times N_{\textrm{col}} \times T$ array $\MATpix$. Moreover, the background levels $b_{i,j}$ are also assumed to be unknown and need to be estimated. 

The target detection problem considered in this paper can be seen as a pixel-wise model selection problem where the parameter space associated with each model is different. To be precise, under $\mathcal{M}_0^{(i,j)}$ (resp. $\mathcal{M}_1^{(i,j)}$), $\btheta_{i,j}^{(0)} \in \bTheta_0$ (resp. $\btheta_{i,j}^{(1)} \in \bTheta_1$) where $\bTheta_0=\bbR^+$ and $\bTheta_1=\bbR^+ \times \bbT \times \bTheta_0$. To simplify notations, the unknown parameter vector is noted $\btheta_{i,j}$ in the remainder of the paper when we do not specify whether it is included in $\bTheta_0$ or $\bTheta_1$.
Estimating $\btheta_{i,j}$ is difficult using standard optimization methods since the dimensionality of the parameter vector depends on the underlying model. 
However, this model selection problem can be solved efficiently in a Bayesian framework by 1) performing inference for each pixel in the parameter space $\left\lbrace \left\lbrace 0\right\rbrace \times \bTheta_0 \right\rbrace \bigcup \left\lbrace  \left\lbrace1 \right\rbrace \times \bTheta_1 \right\rbrace$, 2) incorporating relevant additional prior belief (through prior distributions) and 3) using RJ-MCMC methods adapted for problems whose finite dimensionality in unknown. 

The next section presents a new Bayesian model for target detection accounting for spatial correlations affecting parameters of neighbouring pixels. 
%%%%%%%%%%%%%%%%%%%%%%%%%%%%%%%%%%%%%%%%%%%%%%%%%%%%%%%%%%%%%%%%%%%%%
%%%%%%%%%%%%%%%%%%%%%%%%%%%%%%%%%%%%%%%%%%%%%%%%%%%%%%%%%%%%%%%%%%%%%
\section{Bayesian Model for collaborative target detection via Markovian Dependencies}
\label{sec:Bay_model}

\subsection{Parameter prior distributions}
\label{subsec:priors}
\subsubsection{Priors for the background levels}
In the absence of a target (i.e., assuming \eqref{eq:model0}), gamma distributions are conjugate priors for $b_{i,j}$. Moreover, it has been shown in \cite{Altmann2016a} that considering such priors in the presence of a target also simplifies the sampling procedure. In contrast to the model in \cite{Altmann2016a} which assumed the background levels of the  $N_{\textrm{row}} \times N_{\textrm{col}}$ pixels to be \emph{a priori} independent, here we specify the background levels prior distribution to reflect the prior belief that background levels exhibit spatial correlations. In particular, due to the spatial organization of images, we expect the values of $b_{i,j}$ to vary smoothly from one pixel to another (as will be illustrated in Section \ref{sec:simulations}). In order to model this behaviour,  we specify $\epsilon_{i,j}$ such that the resulting prior for the background matrix $\bfB$ such that $\left[\bfB\right]_{i,j}=b_{i,j}$ is a hidden gamma-MRF (GMRF) \cite{Dikmen2010} (in a similar fashion to the intensity model in \cite{Altmann2016a}).
More precisely, we introduce an $(N_{\textrm{row}}+1) \times (N_{\textrm{col}}+1)$ auxiliary matrix $\bGam$ with elements $\gamma_{i,j} \in \mathbb{R}^+$ and define a bipartite conditional independence graph between $\bfB$ and $\bGam$ such that each $b_{i,j}$ is connected to four neighbour elements of $\bGam$ and vice-versa. This $1$st order neighbourhood structure is similar to that depicted in Fig. 2 in \cite{Altmann2016a}, where we notice that any given $b_{i,j}$ and $b_{i+1,j}$ are $2$nd order neighbours via $\gamma_{i+1,j}$ and $\gamma_{i+1,j+1}$. We specify a GMRF prior for $\bfB,\bGam$ \cite{Dikmen2010}, and obtain the following joint prior for $\bfB,\bGam$\\
$f(\bfB,\bGam|\nu)$\\
\vspace{-0.5cm}
\begin{eqnarray}
\label{eq:GMRF}
 & = & \dfrac{1}{G(\nu)} \prod_{(i,j) 
\in \mathcal{V}_{\bfB}} b_{i,j}^{\left(\nu-1 \right)} \prod_{(i',j') \in \mathcal{V}_{\bGam}} 
\left(\gamma_{i',j'}\right)^{-\left(\nu+1 \right)}\nonumber\\
 & \times & \prod_{\left((i,j),(i',j')\right) \in \mathcal{E}} \exp 
\left(\dfrac{-\nu b_{i,j}}{4 \gamma_{i',j'}} \right),
\end{eqnarray}
%\begin{eqnarray}
%\label{eq:GMRF}
%f(\bfB,\bGam|\nu) & = & \dfrac{1}{G(\nu)} \prod_{(i,j) 
%\in \mathcal{V}_{\bfB}} b_{i,j}^{\left(\nu-1 \right)}\nonumber\\
%& \times & \prod_{(i',j') \in \mathcal{V}_{\bGam}} 
%\left(\gamma_{i',j'}\right)^{-\left(\nu+1 \right)} \nonumber\\
 %& \times & \prod_{\left((i,j),(i',j')\right) \in \mathcal{E}} \exp 
%\left(\dfrac{-\nu b_{i,j}}{4 \gamma_{i',j'}} \right),
%\end{eqnarray}
where $\mathcal{V}_{\bfB}=\left\lbrace 1,\ldots,N_{\textrm{row}}\right \rbrace \times \left\lbrace 1,\ldots,N_{\textrm{col}}\right \rbrace$, $\mathcal{V}_{\bGam}=\left\lbrace 1,\ldots,N_{\textrm{row}}+1 \right \rbrace \times \left\lbrace 1,\ldots,N_{\textrm{col}}+1 \right \rbrace$, and the edge set $\mathcal{E}$ consists of pairs $\left((i,j),(i',j')\right)$ representing the connection between $b_{i,j}$ and $\gamma_{i',j'}$. In this paper, the notation $x|y$ reads ``$x$ conditioned on the value of $y$'' and $f(x|y)$ denotes the probability distribution function of $x|y$, i.e., its probability density function (if $x|y$ is a continuous variable), its probability mass function (if $x|y$ is discrete) or its mixed density if $x|y$ contains discrete and continuous random variables.
It can be seen from \eqref{eq:GMRF} that 
\begin{subeqnarray}
\label{eq:prior1_b}
\slabel{eq:prior1_b2}
b_{i,j}|\left(\bGam,\nu\right) &\sim & \mathcal{G}\left(\nu, \dfrac{\epsilon_{i,j}(\bGam)}{\nu}\right) \\
\slabel{eq:prior1_b3}
\gamma_{i,j}|\left(\bfB,\nu\right) &\sim & \mathcal{IG}\left(\nu,\nu \xi_{i,j}(\bfB)\right)
\end{subeqnarray}
where $\mathcal{G}\left(x,y\right)$ (resp. $\mathcal{IG}\left(x,y\right)$) denotes the gamma (resp. inverse gamma) distribution with shape parameter $x$ and scale parameter $y$, and with
\begin{eqnarray*}
\epsilon_{i,j}(\bGam) & = & 4\left(\gamma_{i,j}^{-1} + \gamma_{i-1,j}^{-1} + \gamma_{i,j-1}^{-1} + \gamma_{i-1,j-1}^{-1}\right)^{-1}\\
\xi_{i,j}(\bfB) &=& \left(b_{i,j} + b_{i+1,j} + b_{i,j+1} + b_{i+1,j+1}\right)/4.
\end{eqnarray*}

Notice that we denote explicitly the dependence on the value of the regularization parameter $\nu$, which here controls the amount of spatial smoothness enforced by the GMRF. Following an empirical Bayesian approach, the value of $\nu$ remains unspecified and will be adjusted automatically during the inference procedure by maximum marginal likelihood estimation.

\subsubsection{Priors for the target parameters}
To reflect the absence of prior knowledge about the target ranges given $z_{i,j}=1$, we assign each possible target depth the following uniform prior $f(t_{i,j}=t) = \dfrac{1}{T'}, \quad t \in \bbT$, 
%\begin{eqnarray}
%p(t_{i,j}=t) = \dfrac{1}{T'}, \quad t \in \bbT,
%\end{eqnarray} 
where $T'=\textrm{card}(\bbT)$. Note however that this prior can be adapted according to potential prior knowledge about the expected target depth distribution.

Accounting for potential spatial dependencies for the target intensities is more challenging as all pixels do not necessarily contain targets. Thus, considering fixed neighbourhood structures (as for $\bfB$) is not well adapted here. Consequently, we propose the following classical hierarchical model
\begin{subeqnarray}
\label{eq:prior_intensity}
r_{i,j}|\left(\alpha,\beta\right) & \sim & \mathcal{G}\left(\alpha,\beta \right), \quad \forall (i,j) \slabel{eq:prior_intensity_21}\\
\alpha|\left(\alpha_1,\alpha_2\right) & \sim & \mathcal{G}\left(\alpha_1,\alpha_2 \right)\slabel{eq:prior_intensity_22}\\
\beta|\left(\beta_1,\beta_2 \right)&  \sim & \mathcal{IG}\left(\beta_1,\beta_2 \right)\slabel{eq:prior_intensity_23}
\end{subeqnarray}
where $(\alpha_1,\alpha_2)$ and $(\beta_1,\beta_2)$ are fixed parameters set to $(\alpha_1,\alpha_2)=(1.1,1)$ and $(\beta_1,\beta_2)=(1,1)$ to reflect the fact that the target intensities have a high probability to be in $(0,1)$. Indeed, the photon impulse response $g_{0}(\cdot)>0$ estimated during the imaging system calibration can be scaled appropriately using reference targets and acquisition times. Although the model \eqref{eq:prior_intensity} does not capture the spatial dependencies between the target intensities, it translates the prior belief that the potential target intensities share similar statistical properties (through $\alpha$ and $\beta$).

\subsubsection{Priors for the observation models}
Finally, in a similar fashion to the background levels, it is often reasonable to expect the probability of a target to be present in a pixel to be related to the presence of targets in the neighbouring pixels (at least when considering targets larger than the spacing between pixels as considered in Section \ref{sec:simulations}). To encode this prior belief, we attach the $N_{\textrm{row}} \times N_{\textrm{col}}$ detection label matrix $\bfZ$ ($\left[\bfZ\right]_{i,j}=z_{i,j}$) the following Ising model 
\begin{eqnarray}
\label{eq:Potts}
f(\bfZ|c) &=& \dfrac{1}{G(c)}\exp\left[c \phi(\bfZ) \right]
\end{eqnarray} 
where $\phi(\bfZ) = \sum_{i,j} \sum_{(i',j') \in \mathcal{V}_{i,j}} \delta \left(z_{i,j}-z_{i',j'} \right)$, 
%\begin{eqnarray}
%\phi(\bfZ) & = & \sum_{i,j} \sum_{(i',j') \in \mathcal{V}_{i,j}} \delta \left(z_{i,j}-z_{i',j'} \right),
%\end{eqnarray}
$\delta(\cdot)$ denotes the Kronecker delta function, and $\mathcal{V}_{i,j}$ is the set of neighbours of pixel $(i,j)$ (in this paper we consider an 8-neighbour structure). Moreover, $c$ is an hyperparameter that controls the spatial granularity of the Ising model and $G(c)=\sum_{\bfZ \in (0;1)^{N_{\textrm{row}} \times N_{\textrm{col}}}} \exp\left[c \phi(\bfZ) \right]$. In a similar fashion to $\nu$, $c$ remains unspecified and will be adjusted automatically during the inference procedure by maximum marginal likelihood estimation using \cite{Pereyra2014ssp}. 
Due to use of the Ising model \eqref{eq:Potts}, we have easy access only to the probability of a target presence in the pixel $(i,j)$ conditioned on the labels values in the neighbouring pixels, i.e., $p_{i,j}=f(z_{i,j}=1|\bfZ_{\backslash (i,j)})$ where $\bfZ_{\backslash (i,j)}$ denotes the subset of $\bfZ$ whose element $z_{i,j}$ has been removed. Consequently, it is important to mention here that the parameters associated with a pixel $(i,j)$ and those of the pixels in $\mathcal{V}_{i,j}$ will not be updated simultaneously. However, chessboard sampling schemes can be used to update conditionally independent sets of parameters in parallel for a more efficient implementation.

In this Section \ref{subsec:priors}, we defined priors distributions for the unknown model parameters. More precisely, priors promoting spatial correlations were used for the background levels and for the detection labels. A joint hierarchical model was proposed for the reflectivity parameters and uniform priors were used for the unknown target depth. In other words, we do not use potential spatial correlation affecting the depth or the reflectivity profiles. Although such consideration could improve the range/reflectivity estimation and possibly the detection performance (which is the main purpose of the proposed method), the model and method we propose can be applied to analyse complex target, possibly highly spatially non-smooth in terms shape and/or reflectivity profile.   
The next Section derives the joint posterior probability associated with the detection problem considered.
\subsection{Joint posterior distribution}
We can now specify the joint posterior distribution for
$\bfZ,\bTheta=\{\btheta_{i,j}\}_{i,j}$ and $\bPhi=\{\bGam, \alpha, \beta\}$ given the observed waveforms $\MATpix$ and the
value of the spatial regularization parameters $\nu$ and $c$ (recall
that their value will be determined by maximum marginal
likelihood estimation during the inference procedure). Using
Bayes’ theorem, and the prior independence assumptions mentioned above, the joint posterior distribution associated with
the proposed Bayesian model is given by
\begin{eqnarray}
\label{eq:joint_post}
f(\bfZ,\bTheta, \bPhi|\MATpix,\nu,c)
 & \propto & \left[\prod_{i,j} f(\Vpix{i,j}|z_{i,j},\btheta_{i,j})f(\btheta_{i,j}|\bfZ,\bPhi)\right]\nonumber\\
& \times &  f(\bfZ|c)f(\bGam|\nu)f(\alpha)f(\beta),
\end{eqnarray}
using $f(\bPhi|\nu)=f(\bGam|\nu)f(\alpha)f(\beta)$.
%
%$f(\bfZ,\bTheta, \bGam, \alpha,\beta|\MATpix,\nu,c)$
%\begin{eqnarray}
%\label{eq:joint_post}
 %& \propto & \left[\prod_{i,j} f(\Vpix{i,j}|z_{i,j},\btheta_{i,j})f(\btheta_{i,j}|\bfZ,\bGam,\alpha,\beta)\right]\nonumber\\
%& \times &  f(\bfZ|c)f(\bGam|\nu)f(\alpha)f(\beta).
%\end{eqnarray}
Note that for clarity the dependence of all distributions on the known fixed quantities $(\alpha_1,\alpha_2,\beta_1,\beta_2)$ is omitted in \eqref{eq:joint_post} and in the remainder of the paper. 
%For illustration, Fig. \ref{fig:DAG} depicts the directed acyclic graph
%(DAG) summarising the structure proposed Bayesian model
%(recall that	$(\bfB,\bGam)$ have a bipartite neighbourhood structure).
%
%
%\begin{figure}[!ht]
%\centerline{ \xymatrix{
  %*+<0.05in>+[F-]+{(\alpha_1,\alpha_2)} \ar@/^/[d] & *+<0.05in>+[F-]+{(\beta_1,\beta_2)} \ar@/^/[d] & &\\
  %\alpha \ar@/^/[rd] & \beta \ar@/^/[d] & \nu \ar@/^/[d] &\\
%\{t_{i,j}\}_{i,j} \ar@/^/[rd] &  \{r_{i,j}\}_{i,j} \ar@/^/[d]& \{\gamma_{i,j}\}_{i,j}, \{b_{i,j}\}_{i,j} \ar@/^/[ld] & c \ar@/^/[lld]\\
    %&  \{z_{i,j}\}_{i,j} \ar@/^/[d] & &  \\
    %& \MATpix &   & }
%} \caption{Directed acyclic graph representing the proposed hierarchical Bayesian model (fixed quantities appear in boxes).} \label{fig:DAG}
%\end{figure}

\section{Bayesian inference} 
\label{sec:Gibbs}

\subsection{Bayesian estimators}
\label{subsec:Bayesian_estimators}
The Bayesian model defined in Section \ref{sec:Bay_model} specifies the joint posterior density for the unknown parameters $\bfZ,\bTheta, \bGam, \alpha$ and $\beta$ given the observed data $\MATpix$ and the parameters $\nu$ and $c$. This posterior distribution models our complete knowledge about the unknowns given the observed data and the prior information available. In this section we define suitable Bayesian estimators to summarize this knowledge and perform target detection. Here we consider the following coupled Bayesian estimators
that are particularly suitable for model selection problems:
the marginal maximum a posteriori (MMAP) estimator for the target presence labels
\begin{eqnarray}\label{zEstimator}
z^{MMAP}_{i,j} = \argmax_{z_{i,j} \in \{0,1\}} f (z_{i,j} | \MATpix, \hat{\nu}, \hat{c}),
\end{eqnarray}
and, conditionally on the estimated labels, 1) the minimum mean square error estimator of the background levels
\begin{eqnarray}\label{bEstimator}
b^{MMSE}_{i,j} = \textrm{E}\left[ b_{i,j} | z_{i,j} = \hat{z}^{MMAP}_{i,j},  \MATpix, \hat{\nu}, \hat{c} \right],
\end{eqnarray}
and 2) for the pixels for which $\hat{z}^{MMAP}_{i,j}=1$, the minimum mean square error estimator of the target intensities
\begin{eqnarray}\label{rEstimator}
r^{MMSE}_{i,j} = \textrm{E}\left[ r_{i,j} | z_{i,j} = 1,  \MATpix, \hat{\nu}, \hat{c} \right],
\end{eqnarray}
and the marginal maximum a posteriori (MMAP) estimator of the target positions
\begin{eqnarray}\label{tEstimator}
t^{MMAP}_{i,j} = \argmax_{t_{i,j} \in \bbT} f (t_{i,j} | z_{i,j} = 1, \MATpix, \hat{\nu}, \hat{c}),
\end{eqnarray}
Note that marginalising out the other unknowns (including $\alpha$ and $\beta$) in \eqref{zEstimator}, \eqref{bEstimator}, \eqref{rEstimator} and \eqref{tEstimator} automatically takes into account their uncertainty.

Computing \eqref{zEstimator} to \eqref{tEstimator} is challenging because it requires having access to the univariate marginal densities of $z_{i,j}$ and the joint marginal densities of (among others) $(b_{i,j}, z_{i,j})$, which in turn require computing the posterior \eqref{eq:joint_post} and integrating it over a very high-dimensional space. Fortunately, these estimators can be efficiently approximated with arbitrarily large accuracy by Monte Carlo integration. This is why we propose to compute \eqref{zEstimator} to \eqref{tEstimator}  by first using an MCMC method to generate samples asymptotically distributed according to \eqref{eq:joint_post}, and subsequently using these samples to approximate the required marginal probabilities and expectations. Here we propose an RJ-MCMC method to simulate samples from \eqref{eq:joint_post}, as this type of MCMC method is particularly suitable for models involving hidden Markov random fields and parameter spaces of varying of dimensions \cite[Chap. 10,11]{Robert2004}. It is important to mention that a standard Gibbs sampler similar to that studied in \cite{Altmann2015b} could have been considered instead of the proposed RJ-MCMC to sample from \eqref{eq:joint_post}. However, such approach would lead to a prohibitively slow exploration of the posterior, mainly due to the weakly informative priors for the target ranges. Precisely, updating sequentially the presence labels and the models parameters generally leads to low probabilities to move from $\mathcal{M}_0^{(i,j)}$ to $\mathcal{M}_1^{(i,j)}$ at each iteration of the sampler. The output of this algorithm are Markov chains of $N_{\textrm{MC}}$ samples distributed according to the posterior distribution \eqref{eq:joint_post}. The first $N_{\textrm{bi}}$ samples of these chains correspond to the so-called \emph{burn-in} transient period and should be discarded (the length of this period can be assessed visually from the chain plots or by computing convergence tests). The remaining  $N_{\textrm{MC}} - N_{\textrm{bi}}$ of each chain are used to approximate the Bayesian estimators \eqref{zEstimator} to \eqref{tEstimator} as in \cite{Altmann2015b,Altmann2016a}. 

Notice that in \eqref{zEstimator} to \eqref{tEstimator}, we have set $c = \hat{c}$ and $\nu = \hat{\nu}$, which denotes the maximum marginal likelihood estimator of the MRF regularisation parameters $c$ and $\nu$ given the observed data $\MATpix$, i.e.,
\begin{eqnarray}\label{alphaML}
(\hat{c},\hat{\nu})=\underset{c \in \mathbb{R}^+, \nu \in \mathbb{R}^+}{\textrm{argmax}} f\left(\MATpix | c,\nu\right),
\end{eqnarray}
This approach for specifying $(c,\nu)$ is taken from the empirical Bayes framework in which parameters with unknown values are replaced by point estimates computed from observed data (as opposed to being fixed \emph{a priori} or integrated out of the model by marginalization). As explained in \cite{Pereyra2014ssp}, this strategy has several important advantages for MRF parameters with intractable conditional distributions such as $c$. In particular, it allows for the automatic adjustment of the value of $(c,\nu)$ for each data set (thus producing significantly better estimation results than using a single fixed value of $(c,\nu)$ for all data sets), and has a computational cost that is several times lower than that of competing approaches, such as including $(c,\nu)$ in the model (by assigning them prior distributions) and subsequently marginalising them during the inference procedure \cite{Pereyra2013ip}.

It is important to mention that the target detection problem considered here and formulated as a model selection problem can be addressed in an optimization framework, e.g., using expectation-maximization methods \cite[Chap. 9]{Bishop2006}. However, in order to assess the relevance of the proposed Bayesian model while minimizing potential algorithmic convergence issues, this paper focuses on a single simulation-based algorithm. Alternative estimation strategies based on the proposed model would deserve a detailed analysis, which is out of scope of this paper.

\subsection{Sampling strategy}
The remainder of this section provides details about the main steps of the proposed sampling strategy, summarised in Algo. \ref{algo:algo2} below. 
\begin{algogo}{Collaborative target detection}
     \label{algo:algo2}
     \begin{algorithmic}[1]
        \STATE \underline{Fixed input parameters:} Lidar impulse response $g_0(\cdot)$, $(\alpha_1,\alpha_2,\beta_2,\beta_2)$, number of burn-in iterations $N_{\textrm{bi}}$, total number of iterations $N_{\textrm{MC}}$
				\STATE \underline{Initialization ($t=0$)}
        \begin{itemize}
        \item Set $\bfZ^{(0)},\bTheta^{(0)},\bGam^{(0)},\alpha^{(0)},\beta^{(0)},\nu^{(0)},c^{(0)}$
        \end{itemize}
        \STATE \underline{Iterations ($1 \leq t \leq N_{\textrm{MC}}$)}
        \STATE Sample $\beta^{(t)} \sim f(\beta|\bfZ^{(t)}, \bTheta^{(t)},\alpha^{(t)})$ in \eqref{eq:post_beta}
				\STATE Sample $\alpha^{(t)} \sim f(\alpha|\bfZ^{(t)}, \bTheta^{(t)},\beta^{(t-1)})$ in \eqref{eq:prior_intensity_22} or \eqref{eq:post_alpha} 
				\STATE Sample $\bGam^{(t)} \sim f(\bGam|\bfB^{(t)},\nu)$ in \eqref{eq:prior1_b3} 
				\FOR{$i=1:N_{\textrm{row}}$}
				\FOR{$j=1:N_{\textrm{col}}$}
				\STATE Set $\bpsi_{i,j}=[p_{i,j},\nu^{(t-1)},\epsilon_{i,j}(\bGam^{(t-1)}),\alpha^{(t-1)},\beta^{(t-1)}]$, where $p_{i,j}$ is computed from \eqref{eq:Potts}
        \STATE Update $(z_{i,j}^{(t)},\btheta_{i,j}^{(t)})$ using Algo. \ref{algo:algo1}
				\ENDFOR
				\ENDFOR
				\STATE Update $(\nu^{(t)},c^{(t)})$ using \cite{Pereyra2014ssp}.
        \STATE Set $t = t+1$.
\end{algorithmic}
\end{algogo}

Each sampling iteration can be decomposed into $5$ main steps (lines $4, 5, 6, (7-12)$ and $13$ in Algo. \ref{algo:algo2}).
Due to the conjugacy between \eqref{eq:prior_intensity_21} and \eqref{eq:prior_intensity_23}, it can be easily shown that 
$f(\beta|\MATpix, \bfZ, \bTheta,\alpha)=f(\beta|\bfZ, \bTheta,\alpha)$ with 
\begin{eqnarray}
\label{eq:post_beta}
\beta|\left(\bfZ, \bTheta,\alpha\right) \sim \mathcal{IG}\left(\beta_1+\alpha\norm{\bfZ}_0,\beta_2+\sum_{z_{i,j}=1}r_{i,j} \right)
\end{eqnarray}
where $\norm{\bfZ}_0$ denotes the number of non-zero elements in $\bfZ$. In a similar fashion to \cite{Zhou2011beta}, when $\norm{\bfZ}_0=0$, i.e., the image does not contain any target, sampling $\alpha$ from its conditional distribution reduces to sampling from \eqref{eq:prior_intensity_22}. If $\norm{\bfZ}_0>0$, 
\begin{eqnarray}
\label{eq:post_alpha}
f(\alpha|\bfZ, \bTheta,\beta) \propto \alpha^{\alpha_1-1}\exp\left(-\frac{\alpha}{\alpha_2}\right)\prod_{(i,j),z_{i,j}=1}\dfrac{r_{i,j}^{\alpha-1}}{\Gamma(\alpha)\beta^\alpha}
\end{eqnarray}
which is strictly log-concave if $\alpha_1 \geq 1$. Consequently, $\alpha$ can be updated using standard Metropolis-Hastings (MH) updates or adaptive rejection sampling \cite{Gilks1992}. Here we used MH updates using a Gaussian random walk whose variance is adjusted during the early iterations of the sampler. By noting that $\bGam$ does not appear in \eqref{eq:model0} nor \eqref{eq:model1}, sampling from its conditional distribution reduces to sampling from $f(\bGam|\bfB,\nu)$ in \eqref{eq:prior1_b3}.

\subsection{RJ-MCMC updates}
To update $(z_{i,j},\btheta_{i,j})$, we construct a transition kernel which admits \\
$f(z_{i,j},\btheta_{i,j}|\Vpix{i,j},\bpsi_{i,j})=$
\begin{eqnarray}
\label{eq:joint_post_RJMCMC}
 f_0(z_{i,j},\btheta_{i,j}|\Vpix{i,j},\bpsi_{i,j}) + f_1 (z_{i,j},\btheta_{i,j}|\Vpix{i,j},\bpsi_{i,j}) 
\end{eqnarray}
as invariant distribution, where\\
$f_0(z_{i,j},\btheta_{i,j}|\Vpix{i,j},\bpsi_{i,j})= \Indicfun{\{0\}\times\bTheta_0}{z_{i,j},\btheta_{i,j}}$
\begin{eqnarray}
 \times \dfrac{f(\Vpix{i,j}|z_{i,j}=0, \btheta_{i,j})f_0(\btheta_{i,j}|\bpsi_{i,j})(1-p_{i,j})}{f(\Vpix{i,j})},
\end{eqnarray}
$f_1(z_{i,j},\btheta_{i,j}|\Vpix{i,j},\bpsi_{i,j})=\Indicfun{\{1\}\times\bTheta_1}{z_{i,j},\btheta_{i,j}}$
\begin{eqnarray}
\times \dfrac{f(\Vpix{i,j}|z_{i,j}=1, \btheta_{i,j})f_1(\btheta_{i,j}|\bpsi_{i,j})p_{i,j}}{f(\Vpix{i,j})},
\end{eqnarray}
with $\bpsi_{i,j}=[p_{i,j},\nu,\epsilon_{i,j},\alpha,\beta]$,
\begin{eqnarray}
f_0(\btheta_{i,j}|\bpsi_{i,j}) &= &f(b_{i,j}|\nu,\epsilon_{i,j}),\nonumber\\
f_1(\btheta_{i,j}|\bpsi_{i,j}) &= &f(b_{i,j}|\nu,\epsilon_{i,j})f(r_{i,j}|\alpha,\beta)f(t_{i,j}=t),\nonumber
\end{eqnarray}
and where $\Indicfun{\mathcal{X}}{\cdot}$ denotes the indicator function defined on $\mathcal{X}$. 
The proposed update scheme, summarized in Algo. \ref{algo:algo1}, can be decomposed into four possible moves detailed in the next few paragraphs. At each iteration, depending on the current state of the chain, two possible moves are proposed. One move proposes a candidate in the current parameter space while the other move proposes a candidate in the other parameter space.
\subsubsection{Fixed dimension moves}

Let $\pi_b>0$ be the probability of proposing a move from model $\mathcal{M}_0^{(i,j)}$ to $\mathcal{M}_1^{(i,j)}$ and $\pi_d>0$ be the probability of proposing the reverse move from model $\mathcal{M}_1^{(i,j)}$ to $\mathcal{M}_0^{(i,j)}$. In this work we set $\pi_b=\pi_d=0.5$.

\begin{algogo}{Reversible Jump MCMC step}
     \label{algo:algo1}
     \begin{algorithmic}[1]
        \STATE \underline{Input parameters:} Current model state $z_{i,j}$ and parameter vector $\btheta_{i,j}$, Hyperparameters $\bpsi_{i,j}$, Observation vector $\Vpix{i,j}$
				\STATE Sample $u \sim \mathcal{U}_{[0,1]}(u)$ and $v \sim \mathcal{U}_{[0,1]}(v)$
				\IF{$z_{i,j}=0$ and $u>\pi_b$}
        \STATE  Set $z_{i,j}^{out}=0$, sample $b_{i,j}$ from \eqref{eq:post_b0} and set $\btheta_{i,j}^{out}=b_{i,j}$
				\ELSIF{$z_{i,j}=0$ and $u\leq\pi_b$} 
				\STATE Sample $(r^*,t^*)$ from \eqref{eq:prop_RJMCMC_2} and set $\btheta_{i,j}^{*(1)}=[r^*,t^*,b_{i,j}]$
				\STATE Compute acceptance ratio $\alpha\left[(0,\btheta_{i,j}^{0}),(1,\btheta_{i,j}^{*1})\right]$ in \eqref{eq:accept_RJMCMC}
				\STATE Sample $v \sim \mathcal{U}_{[0,1]}(v)$
				\IF{$v<\alpha\left[(0,\btheta_{i,j}^{0}),(1,\btheta_{i,j}^{*1})\right]$}
        \STATE  Set $z_{i,j}^{out}=1$, $\btheta_{i,j}^{out}=\btheta_{i,j}^{*1}$
				\ELSE
				\STATE Set $z_{i,j}^{out}=0$, $\btheta_{i,j}^{out}=b_{i,j}$
				\ENDIF
				\ELSIF{$z_{i,j}=0$ and $u>\pi_b$}
        \STATE Set $z_{i,j}^{out}=1$
        \STATE Sample $\btheta_{i,j}^{out}$ from \eqref{eq:post_thet1}
				\ELSE
				\STATE Set $\btheta_{i,j}^{*0}=b_{i,j}$
				\STATE Compute acceptance ratio $\alpha\left[(1,\btheta_{i,j}^{1}),(0,\btheta_{i,j}^{*0})\right]$ in \eqref{eq:accept_RJMCMC_reverse}
				\STATE Sample $v \sim \mathcal{U}_{[0,1]}(v)$
				\IF{$v<\alpha\left[(1,\btheta_{i,j}^{1}),(0,\btheta_{i,j}^{*0})\right]$}
        \STATE  Set $z_{i,j}^{out}=0$, $\btheta_{i,j}^{out}=\btheta_{i,j}^{*0}$
				\ELSE
				\STATE Set $z_{i,j}^{out}=1$, $\btheta_{i,j}^{out}=\btheta_{i,j}$
				\ENDIF
				\ENDIF
        \STATE \underline{Output parameters:} $(z_{i,j}^{out},\btheta_{i,j}^{out})$
        \end{algorithmic}
\end{algogo}

\underline{Move in $\mathcal{M}_0^{(i,j)}$:}
If the current state of the chain is in $\mathcal{M}_0^{(i,j)}$, with probability ($1-\pi_b$), the parameter space remains the same (i.e., $z_{i,j}=0$ remains unchanged) and $\btheta_{i,j}$ is updated according to\\
$b_{i,j}|\left(\Vpix{i,j},z_{i,j}=0,\bpsi_{i,j}\right) \sim$
\begin{eqnarray}
\label{eq:post_b0}
 \mathcal{G}\left(\nu+\norm{\Vpix{i,j}}_1,\left(\frac{\nu}{\epsilon_{i,j}}+T\right)^{-1} \right),
\end{eqnarray}   
using the conjugacy of \eqref{eq:model0} and \eqref{eq:prior1_b2}.\\

\underline{Move in $\mathcal{M}_1^{(i,j)}$:}
Similarly, if the current state of the chain at the $t$th iteration is in $\mathcal{M}_1^{(i,j)}$ (i.e., $\btheta_{i,j}^{(t)}=[r_{i,j}^{(t)},t_{i,j}^{(t)},b_{i,j}^{(t)}]$), with probability ($1-\pi_d$), $\btheta_{i,j}$ is updated using the following three Gibbs sampling steps
\begin{subeqnarray}
\label{eq:post_thet1}
r_{i,j}^{(t+1)} & \sim & f(r_{i,j}|\Vpix{i,j},t_{i,j}^{(t)},b_{i,j}^{(t)},z_{i,j}=1,\bpsi_{i,j}) \slabel{eq:post_r1}\\
t_{i,j}^{(t+1)} & \sim & f(t_{i,j}|\Vpix{i,j},r_{i,j}^{(t+1)},b_{i,j}^{(t)},z_{i,j}=1,\bpsi_{i,j})\slabel{eq:post_t1}\\
b_{i,j}^{(t+1)} & \sim & f(b_{i,j}|\Vpix{i,j},r_{i,j}^{(t+1)},t_{i,j}^{(t+1)},z_{i,j}=1,\bpsi_{i,j}).\slabel{eq:post_b1}
\end{subeqnarray}
It can be shown that sampling from \eqref{eq:post_r1} and \eqref{eq:post_b1} can be achieved by sampling from finite mixtures of gamma distributions and that sampling from \eqref{eq:post_t1} reduces to sampling from a discrete distribution defined on a finite support $\bbT$ (see \cite{Altmann2016a} for details).

\subsubsection{Variable dimension moves}

\underline{Moving from $\mathcal{M}_0^{(i,j)}$ to $\mathcal{M}_1^{(i,j)}$:}
Assume that the current state of the chain is in $\mathcal{M}_0^{(i,j)}$, i.e., $\btheta_{i,j}=\btheta_{i,j}^{0}$. Moving from $\mathcal{M}_0^{(i,j)}$ to $\mathcal{M}_1^{(i,j)}$ requires the construction of an appropriate proposal distribution to propose a candidate $\btheta_{i,j}^{*1} \in \bTheta_1$. Since the background level has the same physical meaning under $\mathcal{M}_0^{(i,j)}$ and $\mathcal{M}_1^{(i,j)}$, it makes sense to use its current value to propose an appropriate candidate and thus increase the acceptance probability of the move (as $\mathcal{M}_0^{(i,j)}$ and $\mathcal{M}_1^{(i,j)}$ and nested models). More precisely, the candidate is constructed as $\btheta_{i,j}^{*1}=[r^*,t^*,\btheta_{i,j}^{0}]$ where $(r^*,t^*)\in \left(\bbR^+ \times \bbT\right)$ is generated according to an arbitrary proposal distribution $q(r^*,t^*|\btheta_{i,j}^{0},\Vpix{i,j},\bpsi_{i,j})$. Assuming that 
\begin{eqnarray}
\label{eq:prop_RJMCMC}
q(r^*,t^*|\btheta_{i,j}^{0},\Vpix{i,j},\bpsi_{i,j})>0,\quad \forall (r^*,t^*)\in \bbR^+ \times \bbT
\end{eqnarray}
leads to the following acceptance ratio
\begin{eqnarray}
\label{eq:accept_RJMCMC}
\alpha\left[(0,\btheta_{i,j}^{0}),(1,\btheta_{i,j}^{*1})\right]=\min \left[1,\rho\left((0,\btheta_{i,j}^{0}),(1,\btheta_{i,j}^{*1})\right)\right]
\end{eqnarray}
where \\
$\rho\left((0,\btheta_{i,j}^{0}),(1,\btheta_{i,j}^{*1})\right)=$
\begin{eqnarray}
\dfrac{f_1(1,\btheta_{i,j}^{*1}|\Vpix{i,j},\bpsi_{i,j})\pi_d}{f_0(0,\btheta_{i,j}^{0}|\Vpix{i,j},\bpsi_{i,j}) q(r^*,t^*|\btheta_{i,j}^{0},\Vpix{i,j},\bpsi_{i,j})\pi_b}.
\end{eqnarray}
Although the choice of the proposal distribution \eqref{eq:prop_RJMCMC} can be arbitrary, it has a significant impact on the acceptance ratio \eqref{eq:accept_RJMCMC} and thus on the mixing properties of the sampler. In this work we use the conditional distribution\\
$f_1(r_{i,j},t_{i,j}|\Vpix{i,j},z_{i,j}=1,b_{i,j},\bpsi_{i,j})$
\begin{eqnarray}
\label{eq:prop_RJMCMC_2}
 \propto f_1(1,\btheta_{i,j}^{1}|\Vpix{i,j},\bpsi_{i,j})
\end{eqnarray} 
as proposal distribution so that the candidate (conditioned on the current value of $b_{i,j}$) lies in a region of relatively high density in $\bTheta_1$. Although the choice of the proposal could be further improved \cite{Brooks2003}, this choice allows frequent moves between $\mathcal{M}_0^{(i,j)}$ and $\mathcal{M}_1^{(i,j)}$ in practice. Note that due to the high posterior correlation of the variables in $\btheta_{i,j}^{1}$ and because $T'=\textrm{card}(\bbT)$ is generally large, using weakly informative proposals instead of \eqref{eq:prop_RJMCMC_2} would lead to prohibitively low acceptance ratios when proposing moves from $\mathcal{M}_0^{(i,j)}$ to $\mathcal{M}_1^{(i,j)}$. Based on \cite{Altmann2016a}, it can be shown that $\forall k \in \bbT$,\\
$f_1(r_{i,j},t_{i,j}=k|\Vpix{i,j},z_{i,j}=1,b_{i,j},\bpsi_{i,j})=$
\begin{eqnarray}
\label{eq:cond_rt}
w_k \sum_{i=0}^{\norm{\Vpix{i,j}}_1} w_{i,k}p_{\mathcal{G}}\left(r_{i,j};\alpha+i,\left(\delta_k + \frac{1}{\beta}\right)^{-1}\right)
\end{eqnarray}
where $\forall k \in \bbT, w_k>0$, $\sum_{k \in \bbT} w_k $, $\forall (k,i), w_{i,k}>0$, $\sum_{i=0}^{\norm{\Vpix{i,j}}_1} w_{i,k}$, 
and $p_{\mathcal{G}}\left(\cdot;\alpha,\beta\right)$ denotes the probability density function of the gamma distribution with shape $\alpha$ and scale $\beta$.  Consequently, sampling from \eqref{eq:prop_RJMCMC_2} can be achieved by first sampling $t_{i,j}$ (using $f_1(t_{i,j}=k|\Vpix{i,j},z_{i,j}=1,b_{i,j},\bpsi_{i,j})=w_k$ by marginalization of $r_{i,j}$) and then sampling $r_{i,j}$ from $f_1(r_{i,j}|\Vpix{i,j},z_{i,j}=1,t_{i,j},\bpsi_{i,j})$. For brevity, the derivation of the probabilities $\{w_k\}$ and $\{w_{i,k}\}$ is omitted here. The interested reader is invited to consult \cite{Altmann2016a} for further details. Note that using \eqref{eq:cond_rt} as a proposal ensures \eqref{eq:prop_RJMCMC} is satisfied and that the generated samples are asymptotically drawn from \eqref{eq:joint_post_RJMCMC} for $(\pi_b,\pi_d) \in (0,1)^2$.

\underline{Moving from $\mathcal{M}_1^{(i,j)}$ to $\mathcal{M}_0^{(i,j)}$:}
Finally, assume now that the current state of the chain is in $\mathcal{M}_1^{(i,j)}$, i.e., $\btheta_{i,j}=\btheta_{i,j}^{1}$.The reverse move associated with the move from $\mathcal{M}_1^{(i,j)}$ to $\mathcal{M}_0^{(i,j)}$ first consists of considering as candidate $\btheta_{i,j}^{*0}=b_{i,j}$, where $b_{i,j}$ is the current value of the background level (under model $\mathcal{M}_1^{(i,j)}$). This candidate (and thus the move to $\mathcal{M}_0^{(i,j)}$) is accepted with probability 
\begin{eqnarray}
\label{eq:accept_RJMCMC_reverse}
\alpha\left[(1,\btheta_{i,j}^{1}),(0,\btheta_{i,j}^{*0})\right]=\min \left[1,\rho\left((1,\btheta_{i,j}^{1}),(0,\btheta_{i,j}^{*0})\right)\right]
\end{eqnarray}
where
$\rho\left((1,\btheta_{i,j}^{1}),(0,\btheta_{i,j}^{*0})\right)= 1/\rho\left((0,\btheta_{i,j}^{*0}),(1,\btheta_{i,j}^{1})\right)$

\section{Simulation results}
\label{sec:simulations}
\subsection{Data acquisition}
\begin{figure}[h!]
  \centering
  \includegraphics[width=0.5\columnwidth]{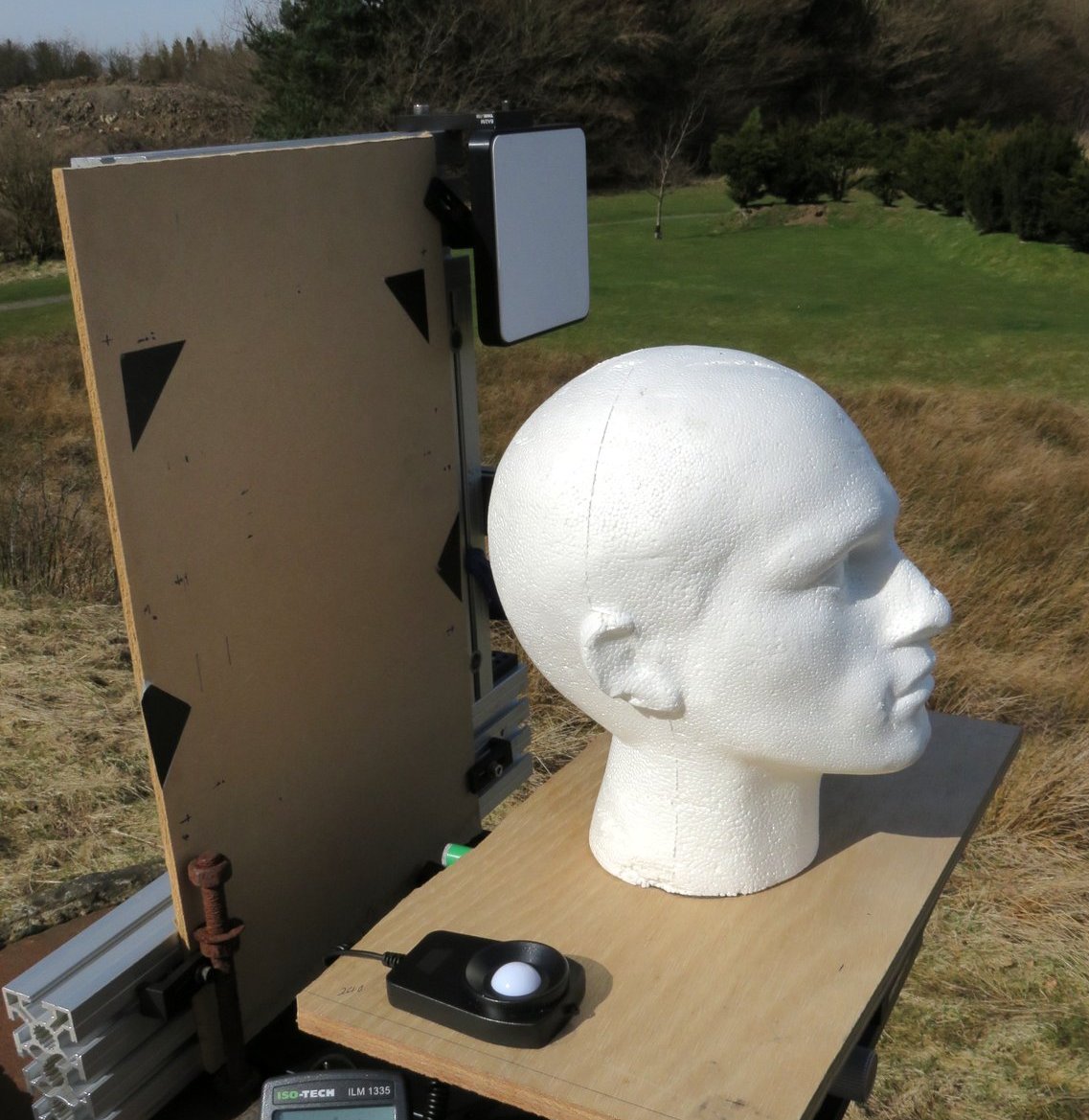}
  \caption{Photograph showing the polystyrene head used for the experiments described here and calibration targets, including
the Spectralon panel (top right corner of the fiberboard).}
  \label{fig:set-up}
\end{figure}
We compare the performance of the method proposed in this paper to reconstruct a depth image of a life-sized polystyrene head (see Fig. \ref{fig:set-up}) located at a distance of $325$ meters from a time-of-flight scanning sensor, based on TCSPC. The transceiver system and data acquisition hardware used for this
work is broadly similar to that described in \cite{McCarthy2009,Krichel2010,Wallace2010,McCarthy2013,Altmann2016a}, and was previously developed at Heriot-Watt University. For the measurements reported in this section, the optical
path of the transceiver was configured to operate with a fiber-coupled illumination wavelength of
841 nm, and a silicon single-photon avalanche diode (SPAD) detector. The overall system had a timing jitter of $60$ps full width at half-maximum (FWHM). The face of the head was pointing towards the scanning system housed in the roof of the lab and a flat medium density fiberboard sheet mounted behind the head target acted as a backboard. This backboard was used for calibration purposes (see Fig. \ref{fig:set-up}) and prevents the detection of return photons from objects placed behind the target of interest, the head in this example. The backboard did not obstruct direct sunlight from illuminating the target, thus not significantly reducing the background levels. To assess the performance of the proposed algorithm, the measured photon histograms have been truncated in order to discard the potential peaks associated with the backplane (i.e., the backplane has been time-gated). The measurements were performed outdoors, on the Edinburgh Campus of Heriot-Watt University, in April 2015 under dry clear skies at $3$ different times of day. That is, the same measurements were repeated at noon, $3$ p.m. and $8$ p.m. (dusk) with atmospheric conditions remaining relatively constant
for the duration of each measurement. The key measurement parameters are summarized in Table \ref{tab:measur_param}. The acquisition time per pixel in Table \ref{tab:measur_param} is $30$ms. However, the data format of time-tagged events allows the construction of photon timing histograms associated with shorter acquisition times, after measurement, as the system records the time of arrival of each detected photon. Here, we evaluate our algorithm for acquisition times of $30$ ms, $3$ ms, $1$ ms, and $300 \mu$s per pixel, for which the number of detected photons per pixels is low. 

\begin{table}[h!]
\renewcommand{\arraystretch}{1.2}
\begin{footnotesize}
\begin{center}
\begin{tabular}{|c|c|}
\hline
Target Stand-off Distance & $\approx 325$m\\
\hline
\multirow{6}{*}{Target Scene} & Polystyrene head \\
& ($\approx 170 \times 285 \times
250$mm\\
& in $W \times H \times D$ when\\
&  viewed from the front)\\
& mounted on a breadboard. \\
& Backplane: MDF board.\\
\hline
\multirow{6}{*}{Laser system} & Supercontinuum \\
& laser source and\\
&  tunable filter \\
& (NKT Photonics)\\
& fiber-coupled to the \\
& custom-designed transceiver unit\\
\hline
Illum. Wavelength & $841$nm\\
\hline
Laser Repetition Rate & $19.5$MHz\\
\hline
Illum. Power at target & $\approx 240\mu$W average optical power\\
\hline
Illum. Beam Diameter at Target & $\approx 10$mm\\
\hline
\multirow{6}{*}{Acquisition Mode} & $200 \times 200$ pixels scan \\
& centred on the head,\\
& covering an area of \\
& $285 \times 285$mm at the scene\\
& Per-pixel acquisition time: $30$ ms\\
& Total scan time: $\approx 20$ minutes\\
\hline
Histogram bin width & $2$ps\\
\hline
Histogram length &  1500 bins (after gating)\\
\hline
Temporal Response of System & $\approx 60$ps FWHM\\
\hline
\end{tabular}
\end{center}
\end{footnotesize}
\caption{Measurement key parameters.\label{tab:measur_param}}
\vspace{-0.3cm}
\end{table}

\begin{figure}[h!]
  \centering
  \includegraphics[width=\columnwidth]{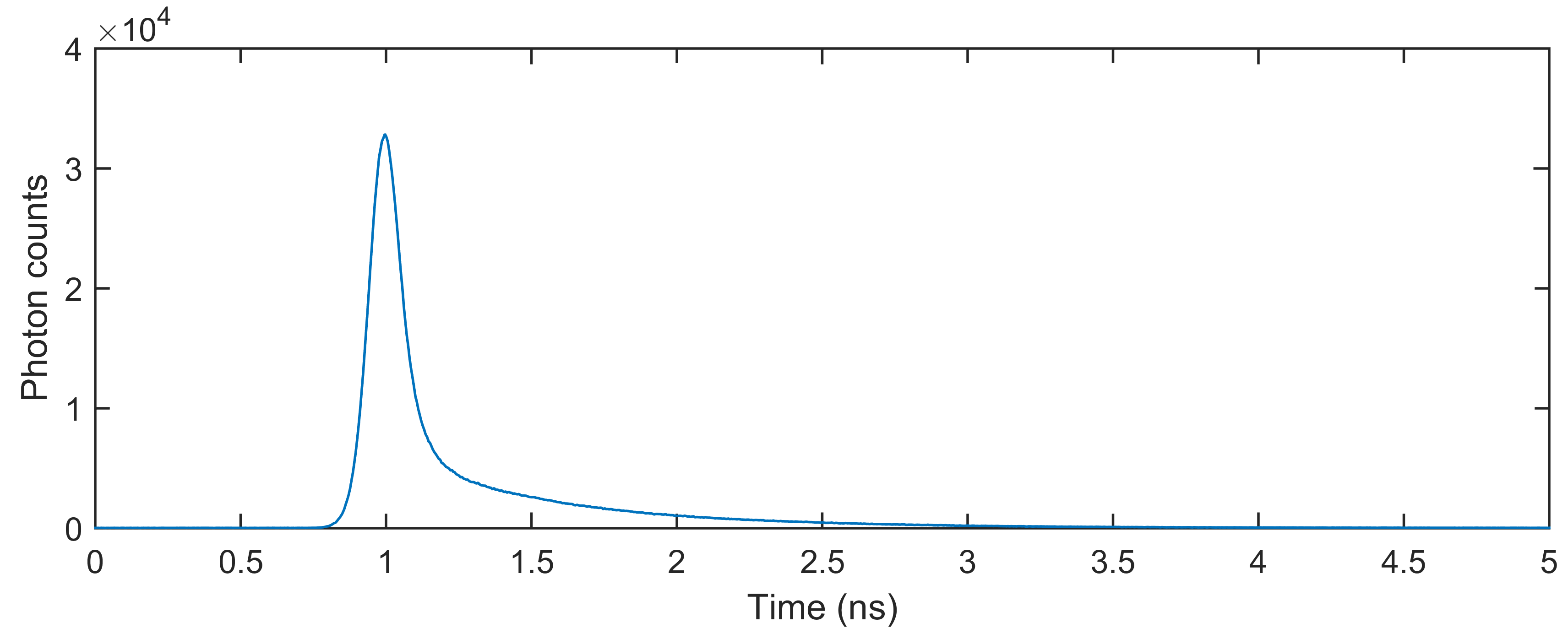}
  \caption{Instrumental response obtained using Spectralon panel placed at $325$m from laser source/detector and for an acquisition time of $100$s (jitter $\approx 60$ps FWHM).}
  \label{fig:impulse_response}
\end{figure}

The instrumental impulse response $g_{0}(\cdot)$ is estimated from preliminary experiments by analysing the distribution of photons reflected onto a Spectralon panel (a commercially available Lambertian scatterer), placed at $325$m from laser source/detector. A long acquisition time ($100$s) is considered here to reduce the impact of photon count variability and a pre-processing step is used to remove the constant background in the measured response. The resulting instrumental impulse response is depicted in Fig. \ref{fig:impulse_response}. Note that the overall shape of this instrumental response, which is the important aspect for depth resolution \cite{Pellegrini2000}, results from a combination of the laser pulse width itself and jitter from a number of sources, including detector jitter (the main contribution in this case) and jitter from the timing electronics (and other electronic components and cabling).

Table \ref{tab:data_analysis} provides details regarding the number of detected photons when varying the acquisition time. The top rows of \ref{tab:data_analysis} show the average number of detected photons per pixel. As expected, the number of detected photons increases linearly with exposure. The bottom rows show that the proportion of empty pixels (no detected photons) increases from noon to $8$ p.m. (due to the average background levels which is higher during the day than at dusk), and that this proportion decreases when increasing the acquisition time (higher probability of detecting photons). Note that for the shortest acquisition ($300\mu$s per pixel), less that $6$ photons per pixel are detected on average, leading to a particularly challenging target detection problem. 
 
\begin{table}[h!]
\renewcommand{\arraystretch}{1.2}
\begin{footnotesize}
\begin{center}
\begin{tabular}{|c|c|c|c|c|c|}
\cline{3-6}
\multicolumn{2}{c|}{} &  \multicolumn{4}{|c|}{Acquisition Time} \\
\cline{3-6}
\multicolumn{2}{c|}{}  & $300\mu$s & $1$ms & $3$ms & $30$ms \\
\hline
\multirow{3}{*}{Av. photon counts} &  noon & $5.6$ & $18.5$ & $55.5$ & $554.6$\\
\cline{2-6}
 &  $3$ p.m. & $4.1$ & $13.7$ & $41.0$ & $408.9$\\
\cline{2-6}
 &  $8$ p.m. & $1.2$ & $4.9$ & $11.6$ & $116.0$\\
\hline
\multirow{3}{*}{Empty pixels ($\%$)} &  noon & $2.79$ & $<0.01$ & $0$ & $0$\\
\cline{2-6}
 &  $3$ p.m. & $4.2$ & $0.02$ & $0$ & $0$\\
\cline{2-6}
 &  $8$ p.m. & $61.8$ & $52.2$ & $40.4$ & $2.2$\\
\hline
\end{tabular}
\end{center}
\end{footnotesize}
\caption{Average number of detected photons per pixel and proportion of empty pixels for the different acquisitions.\label{tab:data_analysis}}
\vspace{-0.3cm}
\end{table}

\subsection{Competing method}
The proposed method is compared to the standard method used for depth imaging \cite{McCarthy2013} and which is divided into two steps. The first step consists of
estimating $t_{i,j}$ using cross-correlation (i.e., by analysing the temporal correlation) between $\log(g_0(\cdot))$ and the photon histogram $\Vpix{i,j}$. For non-empty pixels, i.e., when $\sum_{t=1}^T \pix{i,j}{t}>0$, the object depth is estimated using 
\begin{eqnarray}
\hat{t}_{i,j,\textrm{corr}}=\underset{\tau \in \bbZ}{\textrm{argmax}} \sum_{t=1}^T \pix{i,j}{t} \log\left(g_0(t-\tau)\right).
\end{eqnarray}
which corresponds to log-match filtering or maximum likelihood (ML) estimation under background-free ($b_{i,j}=0$) assumption.
Once the estimated time target distance $\hat{t}_{i,j,\textrm{corr}}$ has been computed, the target intensity and the background level for each pixel are either both set to 0 (for empty pixels) or estimated using ML estimation as 
\begin{eqnarray}
\label{eq:ML_estim}
\left(\hat{r}_{i,j,ML}, \hat{b}_{i,j,ML}\right) = \underset{\substack{r_{i,j}\geq0\\ b_{i,j}\geq0}}{\min} C(\Vpix{i,j},\hat{t}_{i,j,\textrm{corr}},r_{i,j},b_{i,j})
\end{eqnarray} 
with %$C(\Vpix{i,j},t_{i,j},r_{i,j},b_{i,j}) = \sum_{t=1}^{T} y_{i,j,t}\log\left(r_{i,j}g_0\left((t-t_{i,j}\right)+b_{i,j} \right) -\sum_{t=1}^{T} r_{i,j}g_0\left((t-t_{i,j}\right)+b_{i,j}$.
$C(\Vpix{i,j},t_{i,j},r_{i,j},b_{i,j})$
\begin{eqnarray}
\label{eq:cost_function_ML}
 & = & \sum_{t=1}^{T} y_{i,j,t}\log\left(r_{i,j}g_0\left((t-t_{i,j}\right)+b_{i,j} \right)\nonumber\\
 & - & \sum_{t=1}^{T} \left[r_{i,j}g_0\left((t-t_{i,j}\right)+b_{i,j}\right].
\end{eqnarray}
Note that \eqref{eq:cost_function_ML} is convex with respect to $(r_{i,j},b_{i,j})$ with $r_{i,j}\geq 0, b_{i,j}\geq0$ and that \eqref{eq:ML_estim} can be solved efficiently using constrained convex optimization methods (here we used an ADMM method similar to \cite{Figueiredo2010}).  
The proposed Bayesian algorithm has been applied with $N_{\textrm{MC}}=1000$ iterations, including $N_{\textrm{bi}}=300$ burn-in iterations. The computational complexity of the method mainly depends on the average number of detected photons per pixel and the number of admissible target depth $T'$. Indeed, these values have a direct impact on the number of weights to be computed in \eqref{eq:cond_rt}. For a Matlab R2014a implementation on a i7-3.0 GHz desktop computer (16GB RAM), the average computational time is 12 hours per data set, ranging from $1.5$ hour ($8$ p.m., shortest acquisition) to $36$ hours (noon, longest acquisition). It is worth mentioning that other statistical tests, such chi-square tests, could have been considered to perform the target detection, i.e., accept or reject the model \eqref{eq:model0} for each pixel. However, due to the statistical properties of the noise in \eqref{eq:model0} and the low background levels encountered in practice $(b_{i,j}<<1)$, the distribution of the classical chi-square test statistic cannot be accurately approximated by a chi-square distribution and the detection results are generally worse than those obtained with the competing method described in this section.

\subsection{Target detection}
The detection performance of the algorithms is quantitatively assessed by comparing their empirical specificity $\pi_{00}$ (deciding $\mathcal{M}_0^{(i,j)}$ when $\mathcal{M}_0^{(i,j)}$ is true) and sensitivity $\pi_{11}$ or equivalently their empirical probability of false alarm $\pi_{10}=1-\pi_{00}$ and of miss $\pi_{01}=1-\pi_{11}$. Although the standard method does
not provide target detection results directly, it is possible to infer the
target presence by thresholding the estimated intensity images. In
all the results presented here, we set the threshold to $\eta=0.1$, which
corresponds to an estimated target intensity 10 times smaller than
that of the Spectralon panel. Moreover, we observed that this value of threshold yields a satisfactory threshold between high detection performance and high false alarm rate for the data considered. Table \ref{tab:detec_prob} compares the detection performance of the standard and proposed methods, averaged over all pixels, for the $3$ sets of measurements (noon, $3$ p.m. and $8$ p.m.) and for acquisition times of $3$ms, $1$ms and $0.3$ms per pixel. For each data set, the results obtained by the proposed method with an acquisition time of $30$ms are used as ground truth. This table shows that the performance of the two algorithms degrade when reducing the acquisition time. However, the proposed method (as a consequence of its joint detection and estimation ability and the different spatial regularizations) generally provides lower probabilities of false alarm and miss as well as less significant performance degradation than the standard method. 

\begin{table}[h!]
\renewcommand{\arraystretch}{1.2}
\begin{footnotesize}
\begin{center}
\begin{tabular}{|c|c|c|c|c||c|c|}
\cline{4-7}
\multicolumn{3}{c|}{}  &  $\pi_{00}$ & $\pi_{10}$ & $\pi_{01}$ & $\pi_{11}$ \\
\hline
\multirow{6}{*}{8 p.m.} & \multirow{2}{*}{$3$ms}  & X-corr & $98.9$ & $1.1$ & $13.3$ & $86.7$\\
\cline{3-7}
			&	&  Prop. algo. & $\blue{100}$ & $\blue{0}$ & $\blue{9.2}$ & $\blue{90.8}$\\
\cline{2-7}
 & \multirow{2}{*}{$1$ms}  & X-corr & $99.5$ & $0.5$ & $14.1$ & $85.9$\\
\cline{3-7}
			&	&  Prop. algo. & $\blue{100}$ & $\blue{0}$ & $\blue{13.1}$ & $\blue{86.9}$\\
\cline{2-7}
 & \multirow{2}{*}{$0.3$ms}  & X-corr & $99.1$ & $0.9$ & $15.5$ & $84.5$\\
\cline{3-7}
			&	&  Prop. algo. & $\blue{100}$ & $\blue{0}$ & $\blue{8.8}$ & $\blue{91.2}$\\
\hline
\hline
\multirow{6}{*}{3 p.m.} & \multirow{2}{*}{$3$ms}  & X-corr & $86.0$ & $14.0$ & $9.7$ & $90.3$\\
\cline{3-7}
			&	&  Prop. algo. & $\blue{99.9}$ & $\blue{0.01}$ & $\blue{8.8}$ & $\blue{91.2}$\\
\cline{2-7}
 & \multirow{2}{*}{$1$ms}  & X-corr & $60.7$ & $39.3$ & $15.9$ & $84.1$\\
\cline{3-7}
			&	&  Prop. algo. & $\blue{99.9}$ & $\blue{0.01}$ & $\blue{13.9}$ & $\blue{86.1}$\\
\cline{2-7}
 & \multirow{2}{*}{$0.3$ms}  & X-corr & $62.4$ & $37.6$ & $37.3$ & $62.7$\\
\cline{3-7}
			&	&  Prop. algo. & $\blue{99.9}$ & $\blue{0.01}$ & $\blue{15.8}$ & $\blue{84.2}$\\
\hline
\hline
\multirow{6}{*}{noon} & \multirow{2}{*}{$3$ms}  & X-corr & $79.9$ & $20.1$ & $\blue{8.9}$ & $\blue{91.1}$\\
\cline{3-7}
			&	&  Prop. algo. & $\blue{99.9}$ & $\blue{0.01}$ & $10.8$ & $89.2$\\
\cline{2-7}
 & \multirow{2}{*}{$1$ms}  & X-corr & $57.4$ & $42.6$ & $\blue{16.9}$ & $\blue{83.1}$\\
\cline{3-7}
			&	&  Prop. algo. & $\blue{99.9}$ & $\blue{0.01}$ & $18.6$ & $81.4$\\
\cline{2-7}
 & \multirow{2}{*}{$0.3$ms}  & X-corr & $59.6$ & $40.4$ & $39.1$ & $60.9$\\
\cline{3-7}
			&	&  Prop. algo. & $\blue{99.9}$ & $\blue{0.01}$ & $\blue{20.4}$ & $\blue{79.6}$\\
\hline
\end{tabular}
\end{center}
\end{footnotesize}
\caption{Empirical detection performance (probabilities in $\%$, best result in blue).\label{tab:detec_prob}}
\vspace{-0.3cm}
\end{table}

\begin{figure}[h!]
  \centering
  \includegraphics[width=0.5\columnwidth]{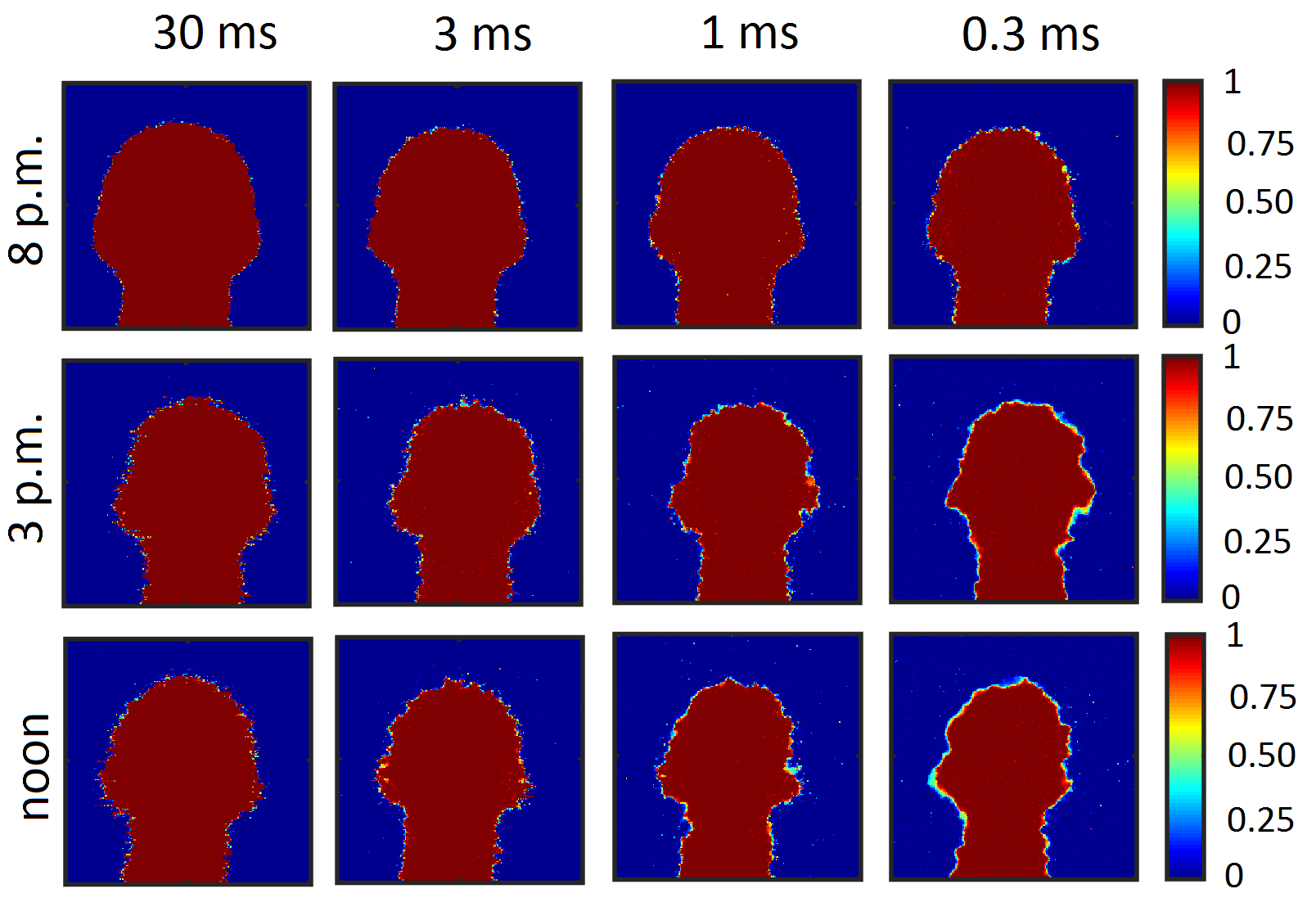}
  \caption{Estimated posterior probability of target presence ($f (z_{i,j}=1 | \MATpix, \hat{\nu}, \hat{c}), \forall (i,j)$) for the different observation conditions considered. Red (resp. blue) regions correspond to high (resp. low) probabilities of presence.}
  \label{fig:post_proba}
	\vspace{-0.3cm}
\end{figure}

For completeness, Fig. \ref{fig:post_proba} depicts the posterior probabilities of target presence, i.e, $f (z_{i,j}=1 | \MATpix, \hat{\nu}, \hat{c}), \forall (i,j)$ for the observations at $8$ p.m. (top), $3$ p.m. (middle) and noon (bottom) and for the different per-pixel acquisition times. This figure shows that the proposed method is able to identify the central region (of high probability) where the head is located and highlights regions of high uncertainty, i.e. where $f (z_{i,j}=1 | \MATpix, \hat{\nu}, \hat{c}) \approx 0.5$, around the boundaries of the head where the detection is difficult due to the low reflectivity of the head is these regions. This figure also shows that the regions of high uncertainty generally broaden as the background levels increase (for fixed target reflectivity) and as the acquisition time decreases.

\subsection{Parameter estimation}
\begin{figure}[h!]
  \centering
  \includegraphics[width=0.5\columnwidth]{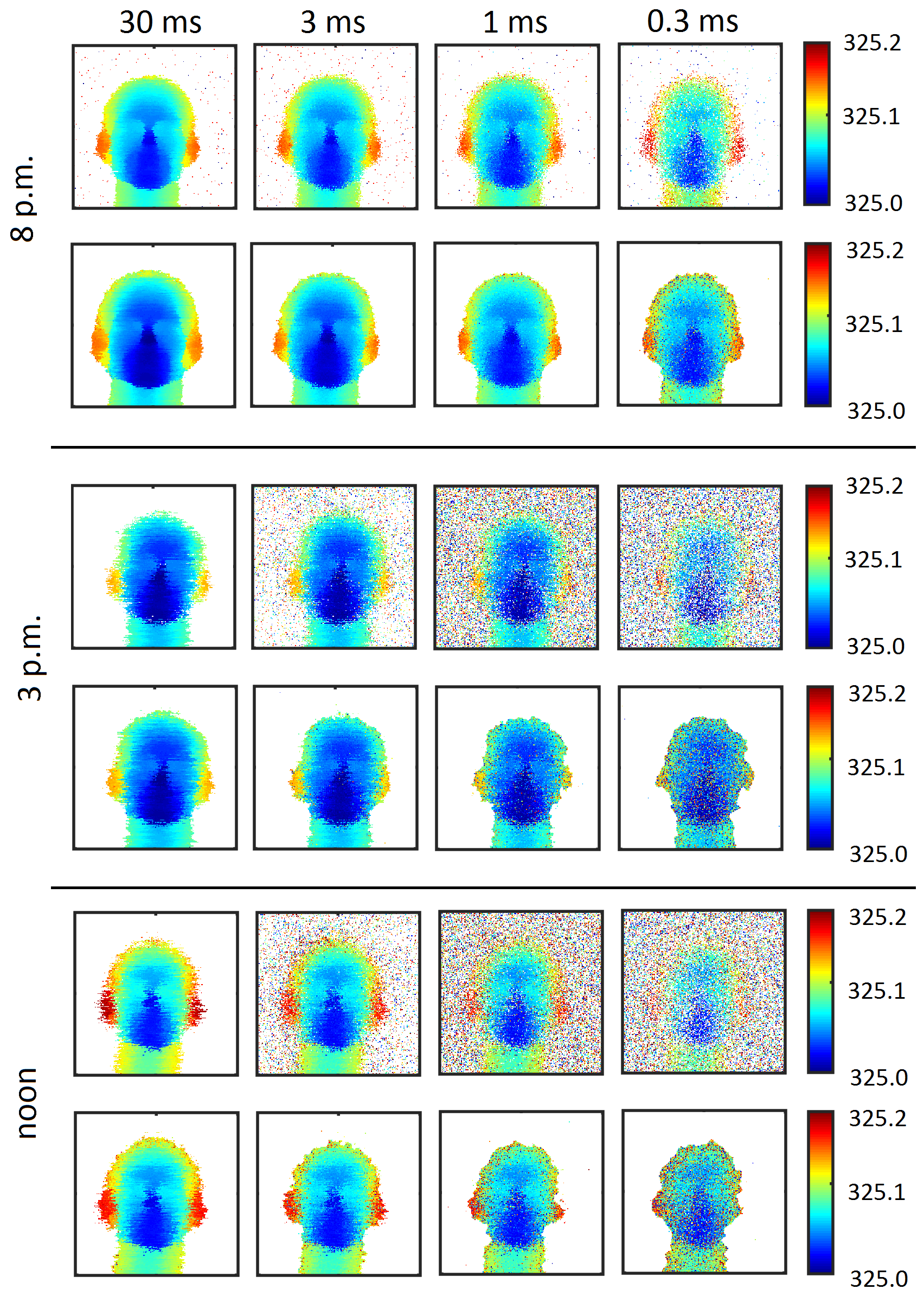}
  \caption{Estimated depths for the $325$m target observed at $8$ p.m. (top), $3$ p.m. (middle) and noon (bottom) and for different per-pixel acquisition times. For each experiment, the top (resp. bottom) row is associated with the standard (resp. proposed) method.}
  \label{fig:depth}
	\vspace{-0.3cm}
\end{figure}

Fig. \ref{fig:depth} compares the estimated depth maps obtained by the standard and the proposed methods. These results show that for large acquisition times, the two methods provide similar results. However, when the acquisition time decreases, the two methods start to fail in identifying the target positions, especially in pixels where no photon is detected. However, due to its better target detection ability, the proposed method provides more reliable depth images as it can more accurately detect pixels not containing a surface.

\begin{figure}[h!]
  \centering
  \includegraphics[width=0.5\columnwidth]{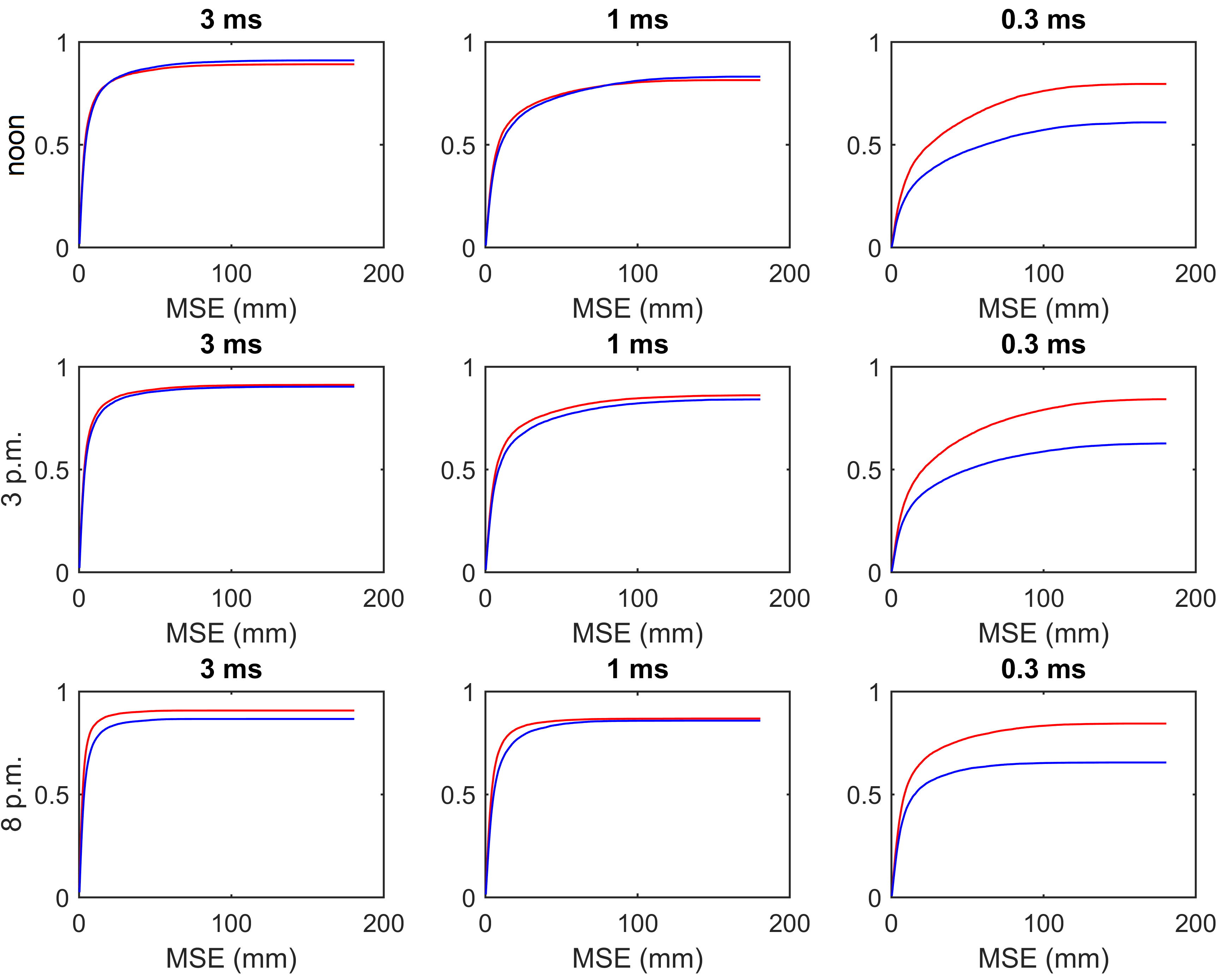}
  \caption{Distance RMSE cdfs provided by the standard (blue) and the proposed (red) methods for the target located at $325$m at observed at noon (top), $3$p.m. (middle) and $8$p.m. (bottom).}
  \label{fig:perf_depth_325m}
	\vspace{-0.3cm}
\end{figure}

The performance of the two methods are quantitatively evaluated using the distance mean squared errors (MSEs) defined by $MSE(d_{i,j})=\norm{\hat{d}_{i,j}-d_{i,j}}_2^2$
where $\norm{\cdots}_2$ denotes the $\ell_2$-norm, $\hat{d}_{i,j}$ is the estimated value of $d_{i,j}=\left(3\times10^{8}\right) t_{i,j}/2$.
Since the actual distances $\{d_{i,j}\}$ are unknown for the data sets considered, these values have been replaced by those estimated by the proposed method for the longest acquisition time ($30$ms). 
Fig. \ref{fig:perf_depth_325m} depict the cumulative density functions (cdfs) of the distance MSEs, defined by $F_d(\tau) = \dfrac{1}{N_{\textrm{row}} N_{\textrm{col}}} \sum_{i,j} \Indicfun{(0,\tau)}{MSE(d_{i,j})}$
%\begin{eqnarray}
%F_d(\tau) & = & \dfrac{1}{N_{\textrm{row}} N_{\textrm{col}}} \sum_{i,j} \Indicfun{(0,\tau)}{MSE(d_{i,j})}
%\end{eqnarray}
where $\Indicfun{(0,\tau)}{\cdot}$ denotes the indicator function defined on $(0,\tau)$. Note that for each dataset, the cdfs are upperbounded by the sensitivity $\pi_{11}$ of each method.
This figure shows that for the pixels containing targets, the two methods provide similar depth estimation performance for the longer acquisition times and that the proposed method is more robust when reducing the acquisition time, thanks to its target detection ability. It is important to recall that in contrast to \cite{Altmann2016a}, the proposed algorithm (as the competing method) does not explicitly account for the spatial correlation of the target depths, which is why the two competing methods provide similar results for long acquisitions. Accounting for such correlations in future target detection and depth imaging methods could further improve the robustness of the method.

\begin{figure}[h!]
  \centering
  \includegraphics[width=0.5\columnwidth]{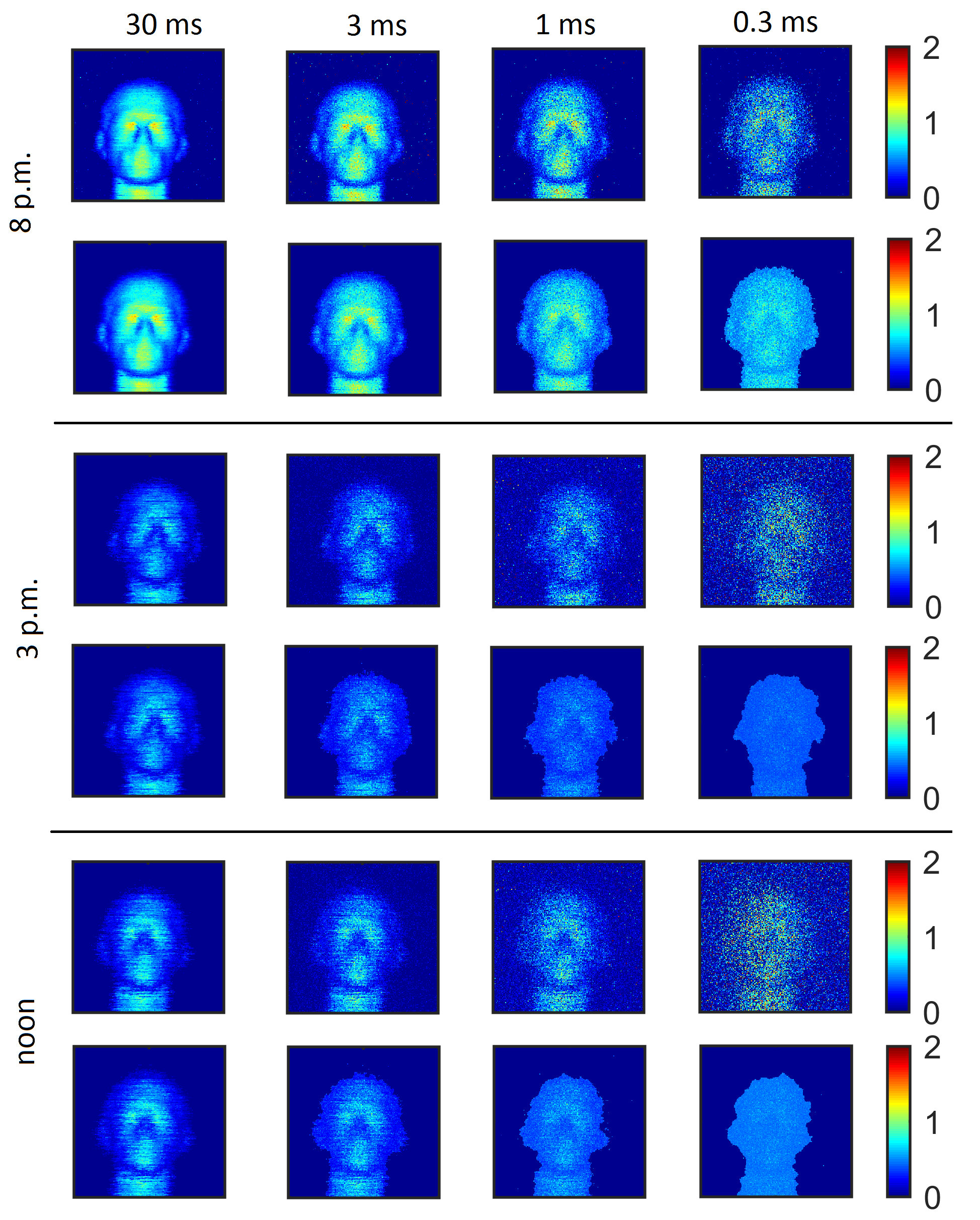}
  \caption{Estimated intensity for the $325$m target observed at $8$ p.m. (top), $3$ p.m. (middle) and noon (bottom) and for different per-pixel acquisition times. For each experiment, the top (resp. bottom) row is associated with the standard (resp. proposed) method.}
  \label{fig:intensity}
	\vspace{-0.3cm}
\end{figure}

Fig. \ref{fig:intensity} compares the estimated intensity maps obtained by the proposed method (for the pixels containing a target) and after thresholding for the standard method ($\eta=0.1$). These results show that the two methods provide similar results for the longest acquisition times and that the proposed method is more robust to the lack/absence of detected photons. In particular, for acquisition times shorter than $3$ms per pixel, it becomes difficult to estimate accurately the intensity of the target, whatever the background levels. By assuming that the target intensities share the same statistical properties (through \eqref{eq:prior_intensity}), the proposed method provides more homogeneous intensity images in the region containing the target than the standard method, which in turn enhance the target detection. Note that for each experiment, the Spectralon response $g_0(\cdot)$ is scaled to account for the acquisition time (e.g., amplitude divided by ten between the $30$ms and $3$ms experiments). 

%
%\begin{figure}[h!]
  %\centering
  %\includegraphics[width=\columnwidth]{intensity_1951.png}
  %\caption{Intensity images for the $325$m target observed at $8$ p.m. and for different per-pixel acquisition times, estimated by the standard method (top row) and the proposed Bayesian algorithm (bottom row).}
  %\label{fig:intensity_1951}
%\end{figure}
%
%\begin{figure}[h!]
  %\centering
  %\includegraphics[width=\columnwidth]{intensity_1508.png}
  %\caption{Intensity images for the $325$m target observed at $3$ p.m. and for different per-pixel acquisition times, estimated by the standard method (top row) and the proposed Bayesian algorithm (bottom row).}
  %\label{fig:intensity_1508}
%\end{figure}
%
%\begin{figure}[h!]
  %\centering
  %\includegraphics[width=\columnwidth]{intensity_1219.png}
  %\caption{Intensity images for the $325$m target observed at $12$ a.m. and for different per-pixel acquisition times, estimated by the standard method (top row) and the proposed Bayesian algorithm (bottom row).}
  %\label{fig:intensity_1219}
%\end{figure}

\begin{figure}[h!]
  \centering
  \includegraphics[width=0.5\columnwidth]{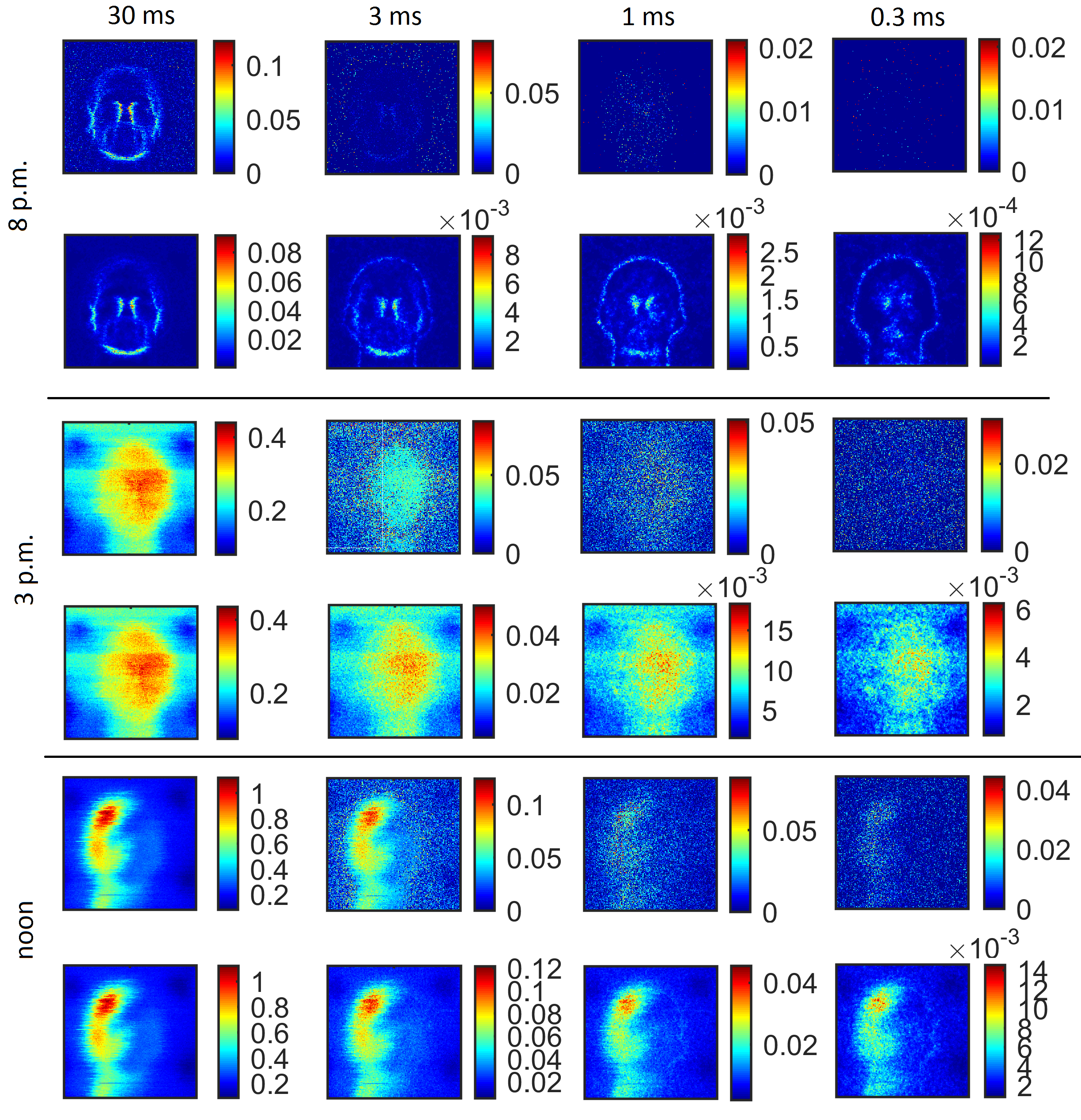}
  \caption{Estimated background levels for the $325$m target observed at $8$ p.m. (top), $3$ p.m. (middle) and noon (bottom) and for different per-pixel acquisition times. For each experiment, the top (resp. bottom) row is associated with the standard (resp. proposed) method.}
  \label{fig:background}
	\vspace{-0.3cm}
\end{figure}

Finally, Fig. \ref{fig:background} compares the background levels estimated by the two methods for the different measurements. The two top rows ($8$ p.m.) of Fig. \ref{fig:background} show that for the longer acquisition times, higher backgrounds are estimated in region of significant depth changes, which can be primarily explained by a model mismatch. In particular, due to the laser beam size and the orientation of the target surface, the peak in the photon histogram can become broader than that depicted in Fig. \ref{fig:impulse_response}. The boundary between the head chin and neck is an even more extreme case where two peaks can be observed. Under brighter observation conditions however (middle and bottom rows of Fig. \ref{fig:background}), these effects become negligible and the background images estimated by the proposed method are in agreement with the observation conditions. Indeed, the detected background photons correspond mainly to photons emitted by external sources (e.g., the sun) and reflected onto the targets. Thus, we can expect (assuming homogeneous ambient illumination) the background levels to be higher in pixels where more reflective surfaces are present. At $3$ p.m. (middle rows), it can be observed that the background levels are generally in the head region and lower in the backplane region. It can be observed that the background levels are particularly low in the regions where black calibration markers have been placed (see Fig. 4 in \cite{Altmann2016a}). Finally, the bottom rows of Fig. \ref{fig:background} clearly show higher background levels on the left-hand side of the head, due to the more direct sun illumination at noon and the head orientation with respect to the sun and observation directions. Note that the regions of particularly low background levels correspond to the four dark triangles mounted on the fiberboard and visible in Fig. \ref{fig:set-up}. 
Fig. \ref{fig:background} also illustrates the benefits or the background model \eqref{eq:prior1_b} used in the proposed method. Accounting for the spatial correlations
of the background levels to regularize the target detection problem 1) provides more realistic background images compared to the pixel-by-pixel method and 2) enhance the target detection. Moreover, by achieving simultaneously the target detection and identification, the proposed method is more robust than the standard method whose performance highly relies on the first depth estimation step.

%\begin{figure}[h!]
  %\centering
  %\includegraphics[width=\columnwidth]{background_1951.png}
  %\caption{Background levels for the $325$m target observed at $8$ p.m. and for different per-pixel acquisition times, estimated by the standard method (top row) and the proposed Bayesian algorithm (bottom row).}
  %\label{fig:background_1951}
%\end{figure}
%
%\begin{figure}[h!]
  %\centering
  %\includegraphics[width=\columnwidth]{background_1508.png}
  %\caption{Background levels for the $325$m target observed at $3$ p.m. and for different per-pixel acquisition times, estimated by the standard method (top row) and the proposed Bayesian algorithm (bottom row).}
  %\label{fig:background_1508}
%\end{figure}
%
%\begin{figure}[h!]
  %\centering
  %\includegraphics[width=\columnwidth]{background_1219.png}
  %\caption{Background levels for the $325$m target observed at $12$ a.m. and for different per-pixel acquisition times, estimated by the standard method (top row) and the proposed Bayesian algorithm (bottom row).}
  %\label{fig:background_1219}
%\end{figure}

\section{Conclusion}
\label{sec:conclusion} 
In this paper, we have presented a Bayesian algorithm for joint target detection and depth imaging using sparse single-photon data. This problem was translated into a pixel-wise model selection problem and a Bayesian hierarchical model was proposed to describe the expected correlations between the pixels of the observed image through appropriate prior distributions. To perform Bayesian inference based on the resulting posterior distribution, we proposed a reversible-jump MCMC algorithm which allows efficient moves between the different parameter spaces. The experiments conducted on real Lidar data demonstrate the ability of the proposed method to 1) detect and 2) identify targets observed under difficult observation conditions (high and spatially variable background levels, short acquisition times), with a better accuracy than existing methods. An important property of the proposed method is its capacity to adjust automatically the different spatial regularization parameters, thus relieving practitioners from the difficult task of setting them by cross-validations. 
In contrast to \cite{Altmann2016a}, we have not explicitly accounted for the possible correlations affecting the intensity and/or depth images. Although it is possible to apply the algorithm studied in \cite{Altmann2016a} to refine the depth/intensity images after the target detection step, e.g., in an empirical Bayes fashion, it would be interesting to extend the model proposed in Section \ref{sec:Bay_model} to capture additional parameter dependency within the target detection procedure. This could be achieved by constructing a depth and/or reflectivity prior model conditioned on the values of the detection labels e.g., $f(\bfT|\bfZ)$. It would also be interesting the correlate the background levels and the target reflectivities. In order to reduced the computational complexity of the target detection, it would be interesting to investigate optimization-based alternatives (e.g., Expectation-
Maximization methods) to be compared with the proposed method in terms of accuracy (estimation performance) and robustness (convergence issues). Finally, the model considered assumed the potential presence of a single target per pixel, which might not be realistic for specific applications. Although the detection of multiple targets using sparse single-photon data is a significantly more difficult problem, extending the model to multiple targets is subject of further investigations that will be reported in subsequent papers.

\bibliographystyle{IEEEtran}
\bibliography{biblio}

\end{document}